\documentstyle[aps,prd,amsfonts]{revtex}

\newcommand{\trK}{\mbox{trK}}
\title{Einstein Constraints on Asymptotically Euclidean Manifolds}
\author{Yvonne Choquet-Bruhat,$^1$ James Isenberg,$^2$
	and James W. York, Jr.$^3$}
\address{$^1$Gravitation et Colmologie Relativiste, t.22-12,
	Un. Paris VI, Paris 75252 France}
\address{$^2$Department of Mathematics, University of Oregon,
	Eugene, OR 97403 USA}
\address{$^3$Department of Physics and Astronomy, University
	of North Carolina, Chapel Hill, NC 27599-3255 USA}

\date{\today}

\begin{document}
\draft
\maketitle

\begin{abstract}
We consider the Einstein constraints on asymptotically euclidean manifolds 
$M$ of dimension $n \geq 3$ with sources of both scaled and unscaled types.
We extend to asymptotically euclidean manifolds the constructive method of
proof of existence.  We also treat discontinuous scaled sources.  In the
last section we obtain new results in the case of non-constant mean curvature.
\end{abstract}

\section{Introduction}
The geometric initial data for the $n+1$ dimensional Einstein equations 
are a properly riemannian metric $\bar{g}$ and a symmetric 2-tensor $K$ on an
$n$-dimensional smooth manifold $M$.  These data must satisfy the constraints,
which are the Gauss-Codazzi equations linking the metric $\bar{g}$ induced on
$M$ by the spacetime metric $g$ with the extrinsic curvature $K$ of $M$ as a
submanifold imbedded in the spacetime $(V,g)$ and the value on $M$ of the 
Ricci tensor of $g$.

As equations on $M$, these constraints read
\begin{eqnarray}
R(\bar{g})-K.K+(\trK)^2&=&2\rho \quad \mbox{hamiltonian constraint} \\
\bar{\nabla}.K-\bar{\nabla}\trK&=&j  \quad \:\mbox{ momentum constraint}
\end{eqnarray}
$R(\bar{g})$ is the scalar curvature and a dot denotes a product 
defined by the metric
$\bar{g}$.  The quantity $\rho$ is a scalar and $j$ a vector on $M$ determined
by the stress energy tensor of the sources.  In coordinates 
adapted to the problem, where the equation of $M$ in $V$ is $x^0=0$, one has

\begin{equation}
j_i=\bar{N}{T_i}^0\; , \quad\rho=\bar{N}^2T^{00}
\end{equation}
with $\rho\geq0$ if the sources satisfy the weak energy condition and if 
$\rho\geq\bar{g}(j,j)^{1/2}$ the sources satisfy the dominant 
energy condition.  The
space scalar $\bar{N}$ is the spacetime lapse function.

A classical method of solving the constraints, initiated by
Lichnerowicz when
$n=3$, is the conformal method (cf.~\cite{cbyo1980} 
and references 
therein anterior to 1980,~\cite{I1995}).  In these papers solutions 
were obtained under the 
condition that the initial submanifold will have constant mean extrinsic 
curvature,
{\it i.e.}, $\trK = \mbox{constant}$.  Recently the results have been 
extended to the 
non-constant mean curvature case with some hypotheses on the smallness of its variations.  
The
case of a compact manifold $M$ is treated in~\cite{cbimo1992} 
and~\cite{imo1996}, 
the first by using the Leray-Schauder theory, the second through a
constructive method.  
Results for asymptotically euclidean $M$ are given in~\cite{C.B.1993}, 
using again the 
Leray-Schauder theory.  All the quoted papers treat the case of scaled and 
continuous sources on a three-dimensional manifold $M$.

We will in this article consider the case where the manifold $M$ has an
arbitrary
dimension $n\geq3$ and the sources are the sum of scaled and unscaled 
ones.  We will
extend to asymptotically euclidean manifolds the constructive method.  
We will
extend the existence proof to discontinuous scaled sources.  

In the last section we obtain results in the non-constant $\trK$ case.  
In the asymptotically euclidean case, non-constant $\trK$ denotes 
non-maximal submanifolds.
A simple smallness assumption on the variations of $\trK$ is sufficient to insure 
existence of solutions for metrics in the positive Yamabe-Brill-Cantor class when 
there are no unscaled sources.  In the other cases the study is more delicate, as 
pointed out by O'Murchadha, and we obtain some results, in particular for unscaled
sources.

We do not claim to have constructed solutions with scaled sources in the negative Yamabe
class on non-maximal manifolds.  The problem of the existence of solutions with large 
variations of $\trK$ remains also open.

We will use the conformal thin sandwich formulation developed recently by one of 
us~\cite{Yo1998} to express the momentum constraint.  It gives a better understanding
of the splitting between given and unknown initial data.

\section{Conformal Method in Its Thin Sandwich Formulation}
\label{ConformalTS}

One turns the hamiltonian constraint into an elliptic equation for a scalar function
$\varphi$ by considering the metric $\bar{g}$ as given up to a conformal factor.  A 
convenient choice is to set, when $n>2$, 
\begin{equation}
\bar{g}\equiv\gamma\varphi^{2p}\; ,\quad\mbox{\it i.e.} \quad \bar{g}_{ij}=
\varphi^{2p}\gamma_{ij}\; ,\quad \mbox{with}\;p=\frac{2}{n-2} \; .
\end{equation}
Then the following identity holds:
\begin{equation}
R(\bar{g})\equiv\varphi^{-(n+2)/(n-2)}\left(\varphi R(\gamma)-\frac{4(n-1)}{n-2}
\triangle_{\gamma}\varphi\right) \; .
\end{equation}
The hamiltonian constraint becomes a semi-linear elliptic equation for $\varphi$ with
a non-linearity of a fairly simple type when $\gamma$ and $K$ are known --- namely
\begin{equation}
\triangle_\gamma\varphi -k_nR(\gamma)\varphi +k_n(K.K-\tau^2+2\rho)\varphi^{(n+2)/(n-2)}=0
\end{equation}
with
\begin{equation}
\tau\equiv\trK \;, \; \; k_n=\frac{n-2}{4(n-1)} \; . 
\end{equation}

We now explain the conformal form of the momentum constraint as recently deduced by
one of us~\cite{Yo1998} from thin sandwich considerations.  It can be construed to include
previous methods as special cases, but no tensor splitting is needed.  The initial metric 
$\bar{g}$ being known up to a 
conformal factor, it is natural to consider that the time derivative of this metric
(the other ingredient of the initial data in a thin sandwich formulation) is known 
only for its conformal equivalence class.  We had above
\begin{equation}
\bar{g}_{ij}=\varphi^{4/(n-2)}\gamma_{ij} \; .
\label{barg}
\end{equation}
If $\bar{g}_{ij}$ and $\gamma_{ij}$ depend on $t$, their time derivatives
are linked by
\begin{equation}
\bar{u}_{ij}=\varphi^{4/(n-2)}u_{ij} \; \mbox{,} \quad \bar{u}^{ij}=\varphi^{-4/(n-2)}u^{ij}
\end{equation}
with
\begin{equation}
\partial_t\bar{g}_{ij}-\frac{1}{n}\bar{g}_{ij}\bar{g}^{hk}\partial_t\bar{g}_{hk}
\equiv\bar{u}_{ij}
\end{equation}
and an analogous expression for $u_{ij}$ constructed with $\gamma_{ij}$.  

We will consider the traceless symmetric two-tensor $u_{ij}$ as given on the manifold
$(M,\gamma)$.  Recall the identity
\begin{equation}
K_{ij}\equiv(2\bar{N})^{-1} \{-\partial_t\bar{g}_{ij}+\bar{\nabla}_i\bar{\beta}_j+
\bar{\nabla}_j\bar{\beta}_i\}  \; \mbox{,}
\label{Kij}
\end{equation}
where $\bar{\beta}$ and $\bar{N}$ will be respectively the shift and the lapse
in the imbedding space time.  The shift vector $\bar{\beta}^i$ is not to be weighted;
it is not a dynamical variable.  The other non-dynamical variable is not the lapse
$\bar{N}$ but a scalar density $\alpha$ of weight $-1$ such that $\bar{N}=\alpha\det
(\bar{g})^{1/2}$
(cf.~\cite{CBYo1995}).  We therefore consider as given in this context a function
$N$ with the space time lapse $\bar{N}$ linked to it by the relation:
\begin{equation}
\bar{N}=\varphi^{2n/(n-2)}N \; .
\label{barN}
\end{equation} 
We denote by $\bar{\nabla}$ and $\nabla$ the covariant derivatives in the metrics $\bar{g}$
and $\gamma$ respectively.  We denote by $\cal L$ the conformal Killing operator
\begin{equation}
(\bar{\cal L}\bar{\beta})^{ij}\equiv \bar{\nabla}^i \bar{\beta}^j + \bar{\nabla}^j \bar{\beta}^i
- \frac{2}{n} \bar{g}^{ij}\bar{\nabla}_h \beta^h .
\end{equation}
We have
\begin{equation}
(\bar{\cal L}\bar{\beta})^{ij}\equiv\varphi^{-4/(n-2)}({\cal L}\beta)^{ij} , \; \; \bar{\beta}^i
\equiv \beta^i
\end{equation}
and
\begin{equation}
K^{ij}=\frac{1}{n}\bar{g}^{ij}\tau+\varphi^{-2(n+2)/(n-2)}A^{ij}
\label{KA}
\end{equation}
with
\begin{equation}
A^{ij}\equiv(2N)^{-1}\{-u^{ij}+({\cal L}\beta)^{ij}\} \; .
\label{Aij}
\end{equation}
One finds by straightforward calculation that the momentum constraint now reads as an
equation on $(M,\gamma)$ with unknown $\beta$ (and $\varphi$ if $D\tau\not\equiv 0)$: 
\begin{eqnarray}
\nabla_j\{(2N^{-1})({\cal L}\beta)^{ij}\}&=&\nabla_j\{(2N^{-1})u^{ij}\} 
+\frac{n-1}{n} \varphi^{2n/n-2}\nabla^i\tau \nonumber\\
& &+\varphi^{2(n+2)/(n-2)}j
\label{NewMomentumConstraint}
\end{eqnarray}
where $N$, $\tau$, and $u$ are given.

The hamiltonian constraint now reads
\begin{eqnarray}
\bigtriangleup_\gamma\varphi-k_nR(\gamma)\varphi+ k_n \varphi^{(-3n+2)/(n-2)}A.A & &\nonumber\\
-\frac{n-2}{4n}\varphi^{(n+2)/(n-2)}\tau^2&=&-2k_n\rho\varphi^{(n+2)/(n-2)}.
\label{NewHamiltonianConstraint}
\end{eqnarray}
The sources are decomposed into scaled and unscaled sources by setting:
\begin{equation}
j\equiv J+\varphi^{-2(n+2)/(n-2)}v \; , \quad \frac{n-2}{2(n-1)}\rho=c+q\varphi^
{-2(n+1)/(n-2)}
\label{ScalarVectorSources}
\end{equation}
More refined decompositions may also occur (See example 2 below).

The energy density scalar $\rho$ and the momentum density vector $j$ behave under conformal
rescaling of the metric according to the source fields which they represent.  See references
by Isenberg, O'Murchadha, and York and by Isenberg and Nester in  \cite{cbyo1980}.

\underline{Examples}.

1.  {\it $n=3$, the source is an electromagnetic or Yang-Mills field $F$}.  The electric and
magnetic fields relative to a spacetime observer at rest with respect to the initial manifold
$M$, ({\it i.e.,} with 4-velocity orthogonal to this manifold), are
\begin{equation}
\bar{E}^i\equiv \bar{N}^{-1} F_0^i = \varphi^{-6}N^{-1}F_0^i \equiv \varphi^{-6} E^i
\end{equation}
\begin{equation}
\bar{H}^i = \frac{1}{2} {\bar{\eta}}^{ijk} F_{jk} = \frac{1}{2} \varphi^{-6} \eta^{ijk}
 F_{jk} \equiv \varphi^{-6} H^i
\end{equation}
with $\eta$ and $\bar{\eta}$ respectively the volume forms of $\gamma$ and $\bar{g}$.

Note that if $(\bar{E}^i, \bar{H}^i)$ satisfy the Maxwell constraints $\bar{\nabla}_i
\bar{E}^i = 0$ and $\bar{\nabla}_i\bar{H}^i = 0$ in the metric $\bar{g}$, the fields 
$(E^i, H^i)$ satisfy these constraints in the metric $\gamma$.  We consider that it is these
last fields which are known on $M$.

The energy density is 
\begin{equation}
\rho = \frac{1}{2} \bar{g}_{ij}(\bar{E}^i \bar{E}^j + \bar{H}^i \bar{H}^j) \equiv \varphi^{-8} q
\end{equation}
with $q$, considered as known on $M$, given by
\begin{equation}
q \equiv \frac{1}{2}\gamma_{ij}(E^i E^j + H^i H^j)
\end{equation}

The momentum density is
\begin{equation}
j^i = \bar{N}T^{i0}= \bar{N}F^{0j}F_j^i = -\bar{E}^j \bar{g}^{ik}\eta_{kjl}H^l = \varphi^{-10}v^i
\end{equation}
with $v^i$ the quantity considered as given;
\begin{equation}
v^i = -\gamma^{ik}\eta_{kjl} E^j H^l \;.
\end{equation}
The sources are scaled as defined and the constraints decouple if $D\tau = 0$.  Note that if
$q \geq (\gamma_{ij} v^i v^j)^{\frac{1}{2}}$, then $\rho \geq (\bar{g}_{ij} 
j^i j^j)^{\frac{1}{2}}$.

2.  {\it General $n$, the source is a Klein-Gordon field.}  The energy density on $M$ of a 
Klein-Gordon field $\psi$ with respect to an observer at rest is
\begin{equation}
\rho = \frac{1}{2}(\bar{N}^{-2}|\partial_0 \psi |^2 + \bar{g}^{ij} \partial_i \psi \partial_j
\psi + m \psi^2) \; ,
\end{equation}
{\it i.e.},
\begin{equation}
\rho = \frac{1}{2} \{\varphi^{\frac{-4n}{n-2}}N^{-2}|\partial_0 \psi |^2 + \varphi^{\frac{-4}{n-2}}
\gamma^{ij}\partial_{i} \psi \partial_j \psi + m \psi^2 \} \;.
\end{equation}
If we consider as known on $M$ the initial data $\psi |_M$ and $\partial_0 \psi |_M$ together
with $\gamma$ and $N$, then neither of the terms in $\rho$ is properly scaled as indicated in (19).  The term 
$N^{-2} | \partial_0 \varphi |^2$ adds in the Hamiltonian constraint to $A.A$, the term
$m^2 \psi$ is unscaled and gives a contribution to $c$, the middle term gives a new, positive,
contribution to the $\varphi$ term which adds to $-R(\gamma)$.  The momentum density is 
\begin{equation}
j^i = -\bar{N}^{-1} \bar{g}^{ij} \partial_j \psi \partial_0 \psi = -\varphi^{-2(n+2)/(n-2)}
\gamma^{ij} \partial_j \psi \partial_0 \psi \;.
\end{equation}
We see that the momentum is properly scaled. {\it The constraints decouple if $D\tau = 0$.}

The methods we give below to study the constraints with properly scaled or unscaled sources 
can be applied to more general scalings, such as this example.

{\bf Summary}.  The given initial data on a manifold $M$ are on the one hand (geometric
initial data) a set $(\gamma, u, \tau, N)$, with $\gamma$ a properly Riemannian metric, 
$u$ a traceless symmetric 2-tensor, $\tau$ and $N$ scalar functions, and on the other hand 
(source data) a set $(J, v, c, q)$, two vectors and two scalars.  The initial data to be
determined by the constraints make a pair $(\varphi, \beta)$ with $\varphi$ a scalar
function and $\beta$ a vector on $M$.  In the conformal thin-sandwich formalism the constraints
reduce to the equation (\ref{NewMomentumConstraint}) and (\ref{NewHamiltonianConstraint})
which read, taking (\ref{ScalarVectorSources}) into account, 
\begin{equation}
\nabla_j \{(2N^{-1})({\cal L}\beta)^{ij}\} = h^i(.,\varphi)
\label{M}
\end{equation}
with
\begin{equation}
h^i(.,\varphi)\equiv\nabla_j\{(2N^{-1})u^{ij}\}+\frac{n-1}{n}\varphi^{2n/(n-2)}
\nabla^i \tau + \varphi^{2(n+2)/(n-2)}J^i+v^i \mbox{,}
\end{equation}
and 
\begin{equation}
\triangle_\gamma\varphi \; = \; f(.,\varphi)\; , \label{H}\\ 
\end{equation}
where
\begin{equation}
f(.,\varphi)\equiv r\varphi-a\varphi^{(-3n+2)/(n-2)} + d\varphi^{(n+2)/(n-2)}
-q\varphi^{-n/(n-2)} \; ,
\label{G}
\end{equation}
with $r$, $a$, and $d$ defined as functions of the geometric data as in equation (45).

When $\tau$ is constant on $M$ and the sources have no unscaled momentum
({\it i.e.} $J=0$) these constraints decouple in the following sense:  the
momentum constraint (\ref{M}) is a linear equation for $\beta$, independent
of $\varphi$, and the Hamiltonian constraint (\ref{H}) is a non-linear equation for 
$\varphi$ when $\beta$ is known.  

When the constraints are solved the spacetime metric reads on $M$:  
\begin{equation}
ds^2 = -\bar{N}^2 dt^2 + \bar{g}_{ij} (dx^i + \beta ^i dt)(dx^j + \beta^j dt) \;,
\end{equation}
with $\bar{g}$ and $\bar{N}$ given by the formulas (\ref{barg}) and (\ref{barN}).
The extrinsic curvature of $M$ is determined by (\ref{KA}) and (\ref{Aij}), the
derivative $\partial_t \bar{g}_{ij}$ on $M$ by (\ref{Kij}).  
The derivatives $\partial_t \bar{N}$ and $\partial_t \beta$
remain arbitrary. 

We now express in our setting the conformal 
invariance of the conformal constraints.

{\bf Lemma} The constraint equations 
(\ref{NewMomentumConstraint}) and
(\ref{NewHamiltonianConstraint}) are 
conformally invariant in the following sense:
If $(\beta,\varphi)$ is a solution of the constraints
with data $(\gamma, u, \tau, N; J, v, c, q)$ then 
$(\beta,\tilde{\varphi})$ is a solution of the
constraints with data 
$(\tilde{\gamma} = (\tilde{\varphi} \varphi^{-1})^{4/(n-2)},
\tilde{u} = (\tilde{\varphi} \varphi^{-1})^{4/(n-2)} u,
\tau, 
\tilde{N} = (\tilde{\varphi} \varphi^{-1})^{2n/(n-2)} N;
\tilde{J} = J, 
\tilde{v} = (\tilde{\varphi} \varphi^{-1})^{-2(n+2)/(n-2)} v,
\tilde{c} = c,
\tilde{q} = (\tilde{\varphi} \varphi^{-1})^{-2(n+1)/(n-2)} q).$

{\it Proof}. If $(\beta, \varphi)$ together with 
the considered given data is a solution of the
conformal constraints, the corresponding Einstein
initial data set $(\bar{g}, K)$ is a solution of
the Einstein constraints with sources $j$, $\rho$
given by (\ref{ScalarVectorSources}).
The Einstein initial data set and sources 
constructed with the $\sim$ quantities
are identical with $(\bar{g}, K)$ and
$(j,\rho)$. Since the Einstein constraints are
satisfied, the conformal constraints written 
with the $\sim$ quantities are also satisfied.

{\it Remark.}  In the case $n=2$, equations analogous to the ones obtained here
for the conformal factor $\varphi$ and the vector $\beta$ are obtained by setting 
(cf.~\cite{Mo1986}):
\begin{equation}
\bar{g}=e^{2\varphi}\gamma \; \; ,
\end{equation}
and in the thin sandwich point of view,
\begin{equation}
\bar{N}=e^{2\varphi}N
\end{equation}
which gives:
\begin{equation}
K^{ij}=e^{4\varphi}A^{ij}+\frac{1}{2}\bar{g}^{ij}\tau \; \; .
\end{equation}
However we will not consider $n=2$  because it poses special problems in what could
correspond to an asymptotically euclidean case.

\section{Asymptotically Euclidean Manifolds and \\Weighted Sobolev Spaces}

The {\it euclidean space} ${\mathbb E}^n$ is the manifold ${\mathbb R}^n$ endowed 
with the euclidean metric which reads in canonical coordinates $\sum(dx^i)^2$.  A
$C^\infty$, $n$-dimensional riemannian manifold $(M,e)$ is called {\it euclidean at 
infinity} if there exists a compact subset $S$ of $M$ such that $M-S$
is the disjoint union of a finite number of open sets $U_i$, and each $(U_i,e)$ is 
isometric to the exterior of a ball in ${\mathbb E}^n$.  Each open set $U_i \subset M$
is sometimes called an {\it end} of $M$.  If $M$ is diffeomorphic to ${\mathbb R}^n$,
it has only one end; and we can then take for $e$ the euclidean metric.  

A riemannian manifold $(M,\gamma)$ is called {\it asymptotically euclidean} if there
exists a riemannian manifold $(M,e)$ euclidean at infinity, and $\gamma$ tends to $e$
at infinity in each end.  Consider one end $U$ and the canonical coordinates $x^i$ in 
the space ${\mathbb E}^n$ which contains the exterior of the ball to which $U$ is 
diffeomorphic.  Set $r\equiv\{\sum(x^i)^2\}^{1/2}$.  In the coordinates $x^i$ the metric
$e$ has components $e_{ij}=\delta_{ij}$.  The metric $\gamma$ tends to $e$ at infinity if
in these coordinates $\gamma_{ij}-\delta_{ij}$ tends to zero.  A possible way of making 
this statement mathematically precise is to use weighted Sobolev spaces.  (One can also
use in these elliptic constraint problems weighted H\"{o}lder spaces, but they are not well
adapted to the related evolution problems). 

A {\it weighted Sobolev space} $W^p_{s,\delta} \; , \; 1\leq p<\infty,s\in
{\mathbb N}_+,\delta \in{\mathbb R}$, of tensors of some given type on the manifold 
$(M,e)$ euclidean at infinity is the closure of $C_0^\infty$ tensors of the given 
type ($C^\infty$ tensors with compact support in $M$) in the norm
\begin{equation}
\| u \|_{W^p_{s,\delta}} = \left\{\sum_{0\leq m\leq s}\int_V \mid \partial^mu 
\mid^p (1+d^2)^{\frac{1}{2} p(\delta+m)}d\mu\right\}^{1/p} \; ,
\end{equation}
where $\partial$, $| \; \; |$ and $d\mu$ denote the covariant derivative, norm and volume 
element in the metric $e$, and $d$ is the distance in the metric $e$ from a point of $M$
to a fixed point. If $(M,e)$ is a euclidean space one can choose $d=r$, the euclidean
distance to the origin.  We recall the multiplication and imbedding properties (cf.~\cite
{CBCh1981})
\begin{eqnarray}
& &W^p_{s_1,\delta_1} \times W^p_{s_2,\delta_2} \subset W^p_{s,\delta} \mbox{ if } s\leq s_1,s_2, 
\; s < s_1 + s_2 - \frac{n}{p} \mbox{ , } \delta < \delta_1 + \delta_2 + \frac{n}{p} 
\mbox{, }\nonumber\\
& &W^p_{s,\delta} \subset C^m_\beta \mbox{  if  } m < s - \frac{n}{p}, 
\; \; \beta <\delta + \frac{n}{p},
 \; \; \nonumber\\
& &\|u\|_{C^m_\beta} \equiv \sum_{0\leq \ell\leq m}\sup_M(|\partial^\ell u|
(1+d^2)^{\frac{1}{2}(\beta+\ell)}).
\end{eqnarray}
The imbedding of the space $W^p_{s,\delta}$ into $W^p_{s^\prime,\delta^\prime}$,
$s\geq s^\prime, \; \delta\geq \delta^\prime$ is compact if $s>s^\prime$, $\delta>
\delta^\prime$.  We have on the other hand:
\begin{equation}
(1+d^2)^{-\beta/2} \in W^p_{s,\delta} \; \mbox{ if } \; \beta>\delta+\frac{n}{p} \; , \quad
s\geq 0 \; .
\end{equation}
Let $(M,e)$ be a manifold euclidean at infinity.  Then the riemannian manifold
$(M,\gamma)$ is said to be ``$W^p_{\sigma,\rho}$ asymptotically euclidean" if 
$\gamma-e \in W^p_{\sigma,\rho}$.  When we speak of ``asymptotically euclidean manifolds" 
without further specification, we suppose that $\gamma-e \in W^p_{\sigma,\rho}$ with
$\sigma>\frac{n}{p}+1$, $\rho>-\frac{n}{p}$. These hypotheses
imply that $\gamma$ is $C^1$ and $\gamma-e$ tends to zero at infinity.

\section{Momentum Constraint}
In the thin sandwich conformal formulation the momentum constraint reads
\begin{equation}
\nabla_j\{(2N^{-1})({\cal L}\beta)^{ij}\}=h(.,\varphi)
\end{equation}
with 
\begin{equation}
h^i(.,\varphi)\equiv\nabla_j\{(2N^{-1})u^{ij}\}+\frac{n-1}{n}\varphi^{2n/(n-2)}
\nabla^i \tau + \varphi^{2(n+2)/(n-2)}J^i+v^i \mbox{,}
\end{equation}
where $N$ and $\tau$ are given functions on $M$ and $u$ a given symmetric
traceless tensor field.  The sources $J$ and $v$ are also considered as known.  We
suppose momentarily that $\varphi$ is also a known function;  in fact, it
disappears from the equation if $\nabla\tau\equiv 0$ and $J\equiv 0$.  

The momentum constraint is a linear elliptic system for the unknown $\beta$ on
the manifold $(M,\gamma)$.  (The symbol of the principal operator is an
isomorphism.)  

{\bf Theorem.}  Let $(M,\gamma)$ be a $W^p_{\sigma,\rho}$ asymptotically 
euclidean manifold with $\sigma>\frac{n}{p}+2, \rho>-\frac{n}{p}$.  Let 
$u$, $\tau \in W^p_{s+1,\delta+1}$ be given, $(1-N^{-1})$ and $(1-\varphi)
\in W^p_{s+2,\delta}, N>0, \varphi>0$, and $J,v\in W^p_{s,\delta+2}$.  The
momentum constraint has one and only one solution $\beta \in W^p_{s+2,\delta}$
if $s>\frac{n}{p}-2$ and $0\leq s\leq\sigma-2, \; -\frac{n}{p}<\delta<n-2-
\frac{n}{p}$.

{\it Proof.}  The operator on the left hand side is injective on
$W^p_{s+2,\delta}$ because a solution $\beta\in W^p_{s,\delta}$, $\delta>
-\frac{n}{p}$, of the equation
\begin{equation}
\nabla_j\{(2N)^{-1}({\cal L}\beta)^{ij}\}=0
\end{equation}
is necessarily a conformal Killing field.  Indeed if $\beta\in C^\infty_0$
the equation implies by integration on $M$ that
\begin{equation}
\int_M\beta_i\nabla_j\left\{(2N)^{-1}({\cal L}\beta)^{ij}\right\}\mu_\gamma
=\int_M(2N)^{-1}{\cal L}\beta.{\cal L}\beta\mu_\gamma=0 \mbox{.}
\end{equation} 
The same is true if $\beta\in W^p_{s+2,\delta^\prime}$ with $\delta^\prime
>-\frac{n}{p}+\frac{n}{2}-2$ (respectively $\delta^\prime\geq -2$ if $p=2$).  
There is such a $\delta^\prime$ if $\beta\in W^p_{s+2,\delta}$ satisfies the 
homogeneous second order equation (cf. a similar proof for the Laplace operator
in the appendix).  It is known that there are no conformal Killing vector
fields tending to zero at infinity on an asymptotically euclidean manifold
(cf.~\cite{cbyo1980})
where a proof requiring only low regularity is cited).

Because the elliptic operator on $\beta$ is injective, the isomorphism theorem
applies to give the existence and uniqueness of $\beta$.

\section{Hamiltonian Constraint}

In the conformal method the hamiltonian constraint reads as a non-linear
elliptic equation for the conformal factor $\varphi$.  We write it
\begin{eqnarray}
\triangle_\gamma\varphi&=&f(.,\varphi) \nonumber\\  \nonumber\\
f(.,\varphi)\equiv r\varphi-a\varphi^{(-3n+2)/(n-2)}&+&d\varphi^{(n+2)/(n-2)}
-q\varphi^{-n/(n-2)} \; ,
\end{eqnarray}
with $A$ given by (\ref{Aij})), and 
\begin{eqnarray}
r &\equiv& k_nR(\gamma), \; a \equiv k_nA.A, \; k_n \equiv (n-2)/4(n-1) \; \nonumber\\
d &\equiv& b-c \; ,\quad b \equiv(n-2)/(4n)\tau^2 \; .
\end{eqnarray}
By their definitions we have
\begin{equation}
a\geq 0, \quad b\geq 0, \quad c\geq 0, \quad q\geq 0.
\end{equation}
The functions $q$ and $c$, scaled and unscaled sources, are considered as given
on $M$.  We will suppose that $\tau$ (hence $b$) is also known on $M$.  The 
function $a$ is known when the momentum constraint has been solved:  this can be 
done independently of $\varphi$ if $\tau$ is constant and the unscaled sources 
have zero momentum.  

The constructive method of sub and super solutions used by one
of us \cite{I1995} to solve non linear elliptic equations
on a compact manifold can be extended to asymptotically
euclidean manifolds.

The following theorem is a particular case of the theorem 
proven in the Appendix B.

{\bf Theorem.}  Let $(M,\gamma)$ be a $(p,\sigma,\rho)$ asymptotically euclidean 
manifold with $\sigma>\frac{n}{p}+1$,  $\rho>-\frac{n}{p}$.  Suppose $r$, $a$,
$q$, $d\in W^p_{s,\delta+2}, \sigma-1\geq s>\frac{n}{p}, -\frac{n}{p}<\delta$.
Suppose the equation $\triangle_\gamma\varphi=f(x,\varphi)$ admits a subsolution 
$\varphi_->0$ and a uniformly bounded supersolution $\varphi_+$, functions in
$C^2$ such that
\begin{equation}
\triangle_\gamma\varphi_-\geq f(.,\varphi_-) \; , \quad \triangle_\gamma
\varphi_+\leq f(.,\varphi_+)
\end{equation}
and
\begin{eqnarray}
\lim_\infty \varphi_-&\leq& 1 \; , \quad \lim_\infty
\varphi_+\geq 1 \nonumber\\
\varphi_-&\leq&\varphi_+ \; \mbox{on} \; M \; .
\end{eqnarray}
Suppose that $D{\varphi_-}, D{\varphi_+} \in W^p_{s^\prime-1,\delta^\prime+1}
\; , \quad s^\prime\geq s \; , \quad \delta^\prime>-\frac{n}{p}$.

Then the equation admits a solution $\varphi$ such that:
\begin{equation}
\varphi_-\leq \varphi\leq \varphi_+ \; , \quad 1-\varphi\in W^p_{s+2,\delta}
\end{equation}
if
\begin{equation}
-\frac{n}{p}<\delta<n-2-\frac{n}{p} \; .
\end{equation}
{\it Remark.}  When $r\equiv k_nR(\gamma)$ we have $r\in 
W^p_{\sigma-2,\delta+2}$ if $\sigma>\frac{n}{p}+2 \; , \; \delta>
-\frac{n}{p}$ .  We will use this theorem directly in Section \ref{GeneralCases},
with constant sub and super solutions.  We will give and use in Sections 
\ref{BrillCantor} and \ref{DiscontinuousSources} intermediate simple steps
to obtain non-constant sub and supersolutions. 

\section{Brill-Cantor Theorem}
\label{BrillCantor}

The constraints in their conformal formulation are invariant under conformal
rescaling (cf. Section \ref{ConformalTS}).
 
In the case of a compact manifold $M$ a convenient first step before 
studying the solution of the Lichnerowicz equation is to use the Yamabe
theorem which says that each manifold $(M,\gamma)$ is conformal to a manifold 
with constant scalar curvature which can be chosen to be $1$, $-1$ or zero.  The 
positive, negative and zero Yamabe classes correspond to the signs of these
constants and are conformal invariants.  There is no known analogous theorem
for asymptotically euclidean manifolds.  (In any case the curvatures could
not be non-zero constants.)  However an interesting theorem has been proved by 
Brill and Cantor, with the following definition.

\underline{Definition.}  The asymptotically euclidean manifold $(M,\gamma)$
is in the positive Yamabe class if for every function $f$ on $M$ with
$f\in C^\infty_0$ , $f\not\equiv 0$, it is true that 
\begin{equation}
\int_M\left\{|Df|^2+r(\gamma)f^2\right\}\mu_\gamma>0 \; .
\end{equation}     
The positive Yamabe class is a conformal invariant due to the identity
\begin{eqnarray}
\triangle_\gamma f - r(\gamma)f&\equiv& \varphi^{(n+2)/(n-2)}\{\triangle_{\gamma^\prime}
f^\prime-r(\gamma^\prime)f^\prime\} \\
\gamma^\prime&=&\varphi^{4/(n-2)}\gamma \; , \quad f^\prime = f\varphi^{-1} \; ,
\end{eqnarray}
which gives after integration by parts with $f \in C^\infty_0$, because 
$\mu_{\gamma^\prime}=\varphi^{2n/n-2}\mu_\gamma$, 
\begin{equation}
\int_M\left\{|Df|^2+r(\gamma)f^2\right\}\mu_\gamma = \int_M\left\{|Df^\prime
|^2+r(\gamma^\prime){f^\prime}^2\right\}\mu_{\gamma^\prime}
\end{equation}

We will say, following O'Murchada that the asymptotically Euclidean manifold $(M,\gamma)$
is in the negative Yamabe class if it is not in the positive one \cite{O'Murchadha1988}.  However, 
analogy with the case
of a compact manifold can be misleading, as shown in the following theorem.

{\bf Theorem.} (\cite{BrillCantor1981}).  The asymptotically euclidean manifold
$(M,\gamma)$ is conformal to a manifold with zero scalar curvature, that is, the
equation $\triangle_\gamma\varphi - r(\gamma)\varphi =0$ has a solution $\varphi >0$,
if and only if $(M,\gamma)$ is in the {\it positive Yamabe class}. 

The physical metric $\bar{g}$ that solves the constraints together with the symmetric 
two-tensor $K$ has a non-negative scalar curvature $R(\bar{g})$ if the sources have 
positive energy and the initial manifold has constant mean extrinsic curvature, (necessarily
zero in the asymptotically euclidean case)..
Thus, $R(\bar{g})\geq 0$, with $R(\bar{g})
\not\equiv 0$ except in vacuum for an instant of time symmetry, {\it i.e.} $K\equiv 0$.
Therefore, the physical metric $\bar{g}$ on an initial maximal submanifold is in the
positive Yamabe class and all metrics $\gamma$ used as substrata to obtain it must
be in that class.

We will prove a more general theorem.  We will also make fewer restrictions than Brill-Cantor 
on the weighted spaces.

{\bf Theorem.}  On a $(p,\sigma,\rho)$ asymptotically euclidean manifold the equation
\begin{equation}
\triangle_\gamma\varphi - \alpha\varphi = \mbox{v} \; ,
\end{equation}
where $\alpha, \mbox{v}\in W^p_{s,\delta+2}, \mbox{v}\leq 0$, has a solution
$\varphi >0, \varphi-1\in W^p_{s+2,\delta}, s\geq 0 \; , \; \delta>-\frac{n}{p}$
only if for all $f\in C^\infty_0 \; , \; f\not\equiv 0$, the following inequality
holds:
\begin{equation}
\int_M\left\{|Df|^2+\alpha f^2\right\}\mu_\gamma>0 \; .
\end{equation}
Under the same hypothesis the solution $\varphi$ exists with $\varphi-1\in W^p_{s+2,\delta}$, 
and $\varphi>0$ if one supposes moreover $s>\frac{n}{p}$, $\delta>\frac{n}{2}- \frac{n}{p} - 1$
if $p\neq 2$ (respectively $\delta \geq -1$ if $p=2$),  and that either 
$\mbox{v}<0$ or $\mbox{v}\equiv 0$ on $M$, or $\alpha=r(\gamma)$ with, in this last case 
$\sigma\geq 2$.

The theorem of Brill and Cantor corresponds to the case $\mbox{v}\equiv 0$ and 
$\alpha=r(\gamma)$.  They made the additional hypothesis $p>n$.

{\it Proof.}\\
1.  (``only if ")  Suppose $\varphi$ exists and solves the equation satisfying the hypothesis
of the theorem.  Then we will show that for any $f\not\equiv 0$, $f\in C^\infty_0$, 
\begin{equation}
\int_M\left\{|Df|^2+\alpha f^2\right\}\mu_\gamma>0 \; .
\end{equation}
Indeed, let $f\in C^\infty_0$, $f\not\equiv 0$.  The function $\theta=f\varphi^{-1}$ has
compact support, belongs to $W^p_{2,\delta^\prime}$ for any $\delta^\prime$ 
and is such that $D\theta\not\equiv 0$ 
since $\theta$, having compact support, cannot be a constant without being identically
zero.  We have by elementary calculus:
\begin{equation}
|Df|^2=|D\theta|^2\varphi^2+\varphi D\varphi.D(\theta^2)+\theta^2|D\varphi|^2 \; .
\end{equation} 
The following integration by parts holds for the considered functions:
\begin{equation}
\int_M\varphi D\varphi.D(\theta^2)\mu_\gamma =\int_M-\theta^2D.(\varphi D\varphi)\mu_\gamma \; ;
\end{equation}
therefore,
\begin{equation}
\int_M\varphi D\varphi .D(\theta^2)\mu_\gamma=\int_M-\theta^2(\varphi\triangle_\gamma\varphi+
|D\varphi|^2)\mu_\gamma
\end{equation}
and
\begin{equation}
\int_M|Df|^2\mu_\gamma>\int_M-\theta^2\varphi\triangle_\gamma\varphi\mu_\gamma \; .
\end{equation}
Hence when $\varphi$ satisfies the given equation and $\theta\varphi =f$:
\begin{equation}
\int_M\left\{|Df|^2+\alpha f^2\right\}\mu_\gamma>\int_M-\mbox{v}\theta^2\varphi\mu_\gamma\geq 0
\end{equation}
if $\varphi>0$ and $\mbox{v}\leq 0$. \bigskip

2.  (``if ":existence)  Setting $\varphi=1+u$ the equation reads:
\begin{equation}
\triangle_\gamma u -\alpha u=\mbox{v}+\alpha \; .
\end{equation}
The operator $\triangle_\gamma-\alpha$ is injective on $W^p_{2,\delta}$ (cf.  Appendix A).

The general theorem on linear elliptic equations on an asymptotically euclidean
manifold shows that our equation has one solution $u\in W^p_{s+2,\delta}$, $s\geq 0$, 
$-\frac{n}{p}<\delta<n-2-\frac{n}{p}$.  The problem is to prove that $\varphi=1+u$ 
is positive.  We will use the maximum principle, supposing the solution to be $C^2$
,{\it i.e.}, $s>\frac{n}{p}$.  Since $\alpha$ is not necessarily positive we cannot
apply directly the maximum principle.  One proceeds as in the Brill-Cantor proof.  
One considers the family of equations, which all satisfy the criterion for the 
existence of a solution $\varphi_\lambda$ with $\varphi_\lambda-1\in W^p_{s+2,\delta}$,
\begin{equation}
\triangle_\gamma\varphi -\lambda\alpha\varphi = \lambda\mbox{v} \; , \quad
\lambda\in [0,1] \; .
\end{equation}
The solutions $\varphi_\lambda$ depend continuously on $\lambda$ and we have
$\varphi_0=1$.  If the function $\varphi_1\equiv \varphi$ takes negative values 
there is one of these functions $\varphi_{\lambda_0}$ which takes positive
or zero values.  The points where $\varphi_{\lambda_0}$ vanishes are minima 
of this function.  It is incompatible with the equation satisfied by $\varphi_{\lambda_0}$
if $\mbox{v}$ is negative at that point.  Therefore we have $\varphi_\lambda>0$
for $\lambda\in [0,1]$ if $\mbox{v}<0$.

To prove that $\varphi_{\lambda_0}>0$, and hence $\varphi_\lambda>0$ for $\lambda\in [0,1]$,
when $\mbox{v}\equiv 0$, we use, as Brill-Cantor, a theorem of Alexandrov: if 
there is a point $x_0$ where $\varphi_{\lambda_0}=0$, it is a minimum of this
function, hence $D\varphi_{\lambda_0}(x_0)=0$.  Since the function $\varphi_{\lambda_0}$
and the function identical to zero take the same value as well as their first
derivatives at $x_0$ and satisfy the same elliptic equation they must coincide
(Alexandrov theorem), a result that contradicts the fact that $\varphi_{\lambda_0}$
tends to 1 at infinity. 

If we know only that $\mbox{v}\leq 0$ but $\alpha=r(\gamma)$ we first conformally transform 
the metric $\gamma$ to a metric $\gamma^\prime=\gamma\psi^{4n/(n-2)}$ with zero
scalar curvature:  this is possible by the previous proof for $\mbox{v}\equiv 0$
(original Brill-Cantor theorem).  The equation to solve is equivalent to the following
equation for $\varphi^\prime =\varphi\psi^{-1}$:
\begin{equation}
\triangle_{\gamma^\prime}\varphi^\prime = \psi^{-(n+2)/(n-2)} \mbox{v}\leq 0 \; .
\end{equation}
whose solution is $\varphi^\prime\geq 1$ because $\varphi^\prime$ cannot attain a 
minimum at a point of $M$ and $\varphi^\prime$ tends to 1 at infinity.

\section{Solution of the Equation  \\[0.25cm]
$\triangle_\gamma\varphi - r(\gamma)\varphi
=b\varphi^{(n+2)/(n-2)}$}  \vspace{.25in}

{\bf Theorem.}  If $b\in W^p_{s,\delta+2}, s>\frac{n}{p}, -\frac{n}{p}<\delta<
n-2-\frac{n}{p}, b\geq 0$, the equation
\begin{equation}
\triangle_\gamma\varphi - r(\gamma)\varphi = b\varphi^{(n+2)/(n-2)}
\end{equation}
on the $(p,\sigma,\rho)$ manifold $(M,\gamma)$, $\sigma>\frac{n}{p}+2,
\rho>\frac{n}{p}$ has a solution $\varphi = 1+u, u\in W^p_{s+2,\delta},
s\leq \sigma , \rho>0$ under one or the other of the following hypotheses:

1.  On the subset of $M$ where $r(\gamma)<0$ there exists a number 
$\mu>0$ such that
\begin{equation}
\sup_{\{x\in M, r(\gamma)(x)<0\}}\frac{|r(\gamma)|}{b}\leq \mu \; .
\end{equation}

2.  $(M,\gamma)$ is in the positive Yamabe class.\\
The solution is unique in both cases.\\
{\it Proof.}

1.  The manifold $(M,\gamma)$ and the function $f(x,y)=r(\gamma)\phi + b
(\phi)^{(n+2)/n-2)}$ satisfy the hypothesis (H) spelled out in Appendix B.  The equation 
admits the subsolution $\varphi_- =0$.  A number $\varphi_+$ is a supersolution if
\begin{equation}
\varphi_+ \geq 1 \mbox{ and } r(\gamma)+b\varphi_+^{4n/(n-2)}\geq 0 \mbox{ on } M \; .
\end{equation}
The second inequality is a consequence of the first if $r(\gamma)\geq 0$.

The hypothesis made on $(M,\gamma)$ on the subset $r(\gamma)<0$ insures the
existence of the number $\varphi_+ \geq \varphi_- \equiv 0$, given by
\begin{equation}
\varphi_+ = \max(1,\mu^{(n-2)/4n}) \; .
\end{equation}
The existence of a solution $\psi$, with $0\leq \psi\leq \varphi_+$ and 
$1-\psi\in W^p_{s+2,\delta}$ results from the general theorem.  Such a solution
can be obtained constructively.  We know that $\psi\not\equiv 0$ since it tends to 1
at infinity.

We show that $\psi>0$ on $M$ by using the Alexandrof theorem as we did in the
proof of the Brill-Cantor theorem:  if $\psi$ vanishes at a point $x_0\in M$ this
point is a minimum of $\psi$, hence $D\psi(x_0)=0$.  The functions $\varphi=\psi$
and $\varphi\equiv 0 $ both satisfy the elliptic equation
\begin{equation}
\triangle_\gamma\varphi - (r(\gamma) + b\psi^{4n/n-2})\varphi = 0 \; .
\end{equation}
They as well as their gradients take the same values, zero, at the point $x_0$,
therefore they coincide.  This contradicts the fact that $\psi$ tends to 1
at infinity, therfore there exists no point $x_0$ where $\psi(x_0)=0$.  Hence
$\psi>0$ on $M$.

2.  If $(M,\gamma)$ is in the positive Yamabe class we conformally transform it to
a manifold $(M,\gamma^\prime)$ such that $r(\gamma^\prime)\equiv 0$.  The 
subset of $M$ where $r(\gamma^\prime)<0$ is empty;  therefore, $\varphi_+=1$ can
be chosen as a supersolution.  The proof that $\varphi>0$ on $M$ can be made
using simply the maximum principle:  a solution $\varphi\in C^2$ of the equation
\begin{equation}
\triangle_\gamma\varphi-b\varphi^{(n+2)/(n-2)}\equiv\triangle_\gamma\varphi-
(b\varphi^{4n/(n-2)})\varphi=0
\end{equation}
with $b\geq 0$ cannot attain a nonpositive minimum on $M$ without being a 
constant (which is not possible with $\varphi$ tending to 1 at infinity except
if $b\equiv 0$, in which case $\varphi\equiv 1$).

The uniqueness property in case 2 is simply a consequence of $b\geq 0$ and of the increasing
property with $\varphi>0$ of the function $\varphi^{(n+2)/(n-2)}$, together
with the fact that the difference of two solutions tends to zero at infinity.
The uniqueness in the general case results from the conformal properties.  Indeed
suppose the equation 
\begin{equation}
\triangle_\gamma \varphi - r(\gamma) \varphi = b\varphi^Q \;, \; Q = \frac{n+2}{n-2}
\end{equation}
has two solution $\varphi_1$ and $\varphi_2$.  We deduce from the conformal identity 
\begin{equation}
\triangle_\gamma \varphi - r(\gamma) \varphi = -r(\bar{g})\varphi^Q \; , \; \bar{g}
= \varphi^{2p} \gamma \; , \; p = \frac{2}{n-2}
\end{equation}
that
\begin{equation}
r(\bar{g}) = r(\gamma \varphi_1^{2p}) = r(\gamma \varphi_2^{2p}) = -b \; .
\end{equation}

Consider the identity
\begin{equation}
\triangle_{\gamma\varphi_1^{2p}} (\varphi_1^{-1} \varphi_2) - r(\gamma \varphi_1^{2p})
\varphi_1^{-1}\varphi_2 \equiv -r(\gamma \varphi_2^{2p}) (\varphi_1^{-1} \varphi_2)^Q \; .
\end{equation} 
It implies, because of the previous equalities,
\begin{equation}
\triangle_{\gamma\varphi_1^{2p}} (u-1)-bu \frac{u^{Q-1}-1}{u-1} (u-1)=0 \; ,
\quad u \equiv \varphi_1^{-1} \varphi_2 \; .
\end{equation}
We have $b \geq 0$, $u > 0$, $\frac{(u^{Q-1}-1)}{(u-1)} >0$ since
$u>0$ and $Q > 1$. We deduce from the fact that $u-1$ tends to 
zero at infinity that $u-1=0$ on $M$, {\it i.e.}, $\varphi_1 \equiv \varphi_2$.

{\it Remark.}  By the above theorem, under the hypothesis made, an
asymptotically euclidean manifold $(M,\gamma)$ is conformal to a metric
$\gamma^\prime$ of given non-positive scalar curvature $r(\gamma^\prime)$,
and the solution $\varphi$ of the equation (66) with $b=-r(\gamma^\prime)$ gives
the conformal factor.  (This result was known to O'Murchadha.)

\section{Solution of the Equation \\[0.25cm]
 $\triangle_\gamma\varphi - r(\gamma)\varphi
+ a\varphi^{-P}+q\varphi^{-P^\prime}=0 \; , \quad a\geq 0 \; , \; q\geq 0$}

This equation is the conformal expression of the Hamiltonian constraint
on a maximal manifold with no unscaled sources. The following theorem has been
proved independently in the case $n=3$ in 1979 by Cantor, and
Chaljub-Simon and Choquet-Bruhat (in weighted Holder spaces).
We give here a new constructive proof;
the corresponding function $f(x,\varphi)$ satisfies the hypothesis H of 
Appendix B on any interval $[\ell,\infty), \; \ell>0$.

The generalized Brill-Cantor theorem shows that the considered equation can
have a solution $\varphi>0$ only if $(M,\gamma)$ is in the positive Yamabe class,
a result in agreement with the fact that the original hamiltonian constraint on an
initial maximal submanifold $(M,\bar{g})$ implies $r(\bar{g})\geq 0$.

{\bf Theorem.}  The equation on the $(p,\sigma,\rho)$ asymptotically euclidean
manifold $(M,\gamma)$, $\sigma > \frac{n}{p}+2 \; , \; \rho > \frac{n}{2}-2-\frac{n}{p}$
if $p\neq 2$, and $\rho\geq -1$ if $p=2$, given by 
\begin{eqnarray}
\triangle_\gamma\varphi - r(\gamma)\varphi &=& -a\varphi^{-P} - q\varphi^{-P^\prime}
\; , \; a\geq 0 \; , \; q\geq 0 \; ,\nonumber\\
P&=&(3n-2)/(n-2) \; , \; P^\prime = n/(n-2) \; ,\nonumber\\
a \; , \; q&\in& W^p_{s,\delta+2}, \quad \sigma-2\geq s>\frac{n}{p} \; , \; 
n-2-\frac{n}{p}>\delta>-\frac{n}{p} \; ,
\end{eqnarray}
has a solution $\varphi>0$, $\varphi-1\in W^p_{s+2,\delta}$ if and only if $(M,\gamma)$ 
is in the positive Yamabe class. This solution is such that $\varphi\geq 1$.  It can be 
obtained constructively and is unique.

{\it Proof.}\\
1.  (``only if")  This part follows from the generalized Brill-Cantor theorem.\\
2.  (``if")  The manifold $(M,\gamma)$ is conformal to a manifold $(M,\gamma^\prime)$
with zero scalar curvature, $\gamma^\prime\equiv \psi^{4/(n-2)}\gamma \; , \;
r(\gamma^\prime)=0$.  Conformal covariance shows that
the resolution of the given equation is therefore equivalent
to the resolution of an equation of the same type but with no linear term, which,
suppressing primes, we write as
\begin{equation}
\triangle_\gamma\varphi=-a \varphi^{-P} - q\varphi^{-P^\prime} \; , \quad
a\geq 0 \; , \; q\geq 0 \; .
\end{equation} 
This equation admits a constant subsolution $\varphi_-=1$ but no finite constant
supersolution.  However, it is possible to construct a sequence $u_{\nu}\in 
W^p_{s+2,\delta}$ starting from the subsolution $\varphi_-=1$ by solving the equations
with $k\geq 0 \; , \; k\in W^p_{s,\delta +2}$:
\begin{equation}
\triangle_\gamma u_{\nu}-ku_{\nu}=-a(1+u_{\nu-1})^{-P} - 
q(1+u_{\nu-1})^{-P^\prime} - ku_{\nu-1}\; .
\end{equation}
We have $u_{\nu}\in W^p_{s+2,\delta}\subset C^2_\alpha$ for all $\alpha$ such that
$\alpha<\delta+\frac{n}{p}$, hence $u_{\nu}$ tends to zero at infinity and we can 
use the maximum principle to see that $u_{\nu}\geq 0$.  We could choose 
$k\geq Pa+P^\prime q$ and deduce as before from the maximum principle that the
sequence $u_{\nu}$ is pointwise increasing, but we do not obtain an upper bound
through the maximum principle because we do not have a supersolution.  We choose first 
instead $k=0$ to construct our sequence and write the elliptic estimate, using the
fact that $(1+u_{\nu})^{-P}\leq 1$ since $u_{\nu}\geq 0 \;$ ,
\begin{equation}
\|u_{\nu}\|_{W^p_{2,\delta}}\leq C \left\{\|a\|_{W^p_{0,\delta + 2}}  +
\|q\|_{W^p_{0,\delta + 2}}\right\} \; .
\end{equation}
The sequence, being uniformly bounded in $W^p_{2,\delta}$, admits a subsequence which 
converges in the $W^p_{1,\delta^\prime}$ norm, $\delta^\prime<\delta$, to an
element $u\in W^p_{2,\delta}$.  The rest of the proof is the same as in the general
arguments given in Appendix B, except that in the present case, the sequence $u_\nu$
is not proven to be monotonic, nor identical to the subsequence which converges.  Hence
we cannot conclude that the limit $u$ of the subsequence is a solution of (78).

To obtain a converging sequence, and consequently a solution, we again use (79), but 
now with $k\geq Pa + P^\prime q$.  For (79) with such a $k$, the subsequence limit $u$ serves
as a supersolution.  Therefore, the increasing sequence $u_\nu$ is bounded above by
$u$ and it converges
to it in $W^p_{s+2,\delta}$.  We have $\varphi \geq 1$.  A pointwise upper bound for
$\varphi$ can be deduced from the $W^p_{s+2,\delta}$ norm of $u=\varphi-1$.

{\it Remark.}  The sequence $u_{\nu}$ and the limit $u$, bounded in 
$W^p_{2,\delta}$ norm in terms of the $W^p_{0,\delta+2}$ norms of $a$ and $q$, are
therefore bounded in $C^0_\alpha$ norm in terms of these norms of $a$ and $q$
{\it if} $p>\frac{n}{2}$.

3.  Uniqueness:  the equation with $r(\gamma)=0$ has a unique solution such that
$\varphi$ tends to 1 at infinity because of the monotonicity of the right hand side and
the maximum principle.  The original equation also has a unique solution.

\section{Solution for Scaled Sources}

We now prove an existence theorem for the non-linear elliptic equation
for $\varphi$ expressing the Hamiltonian constraint on an arbitrary
initial manifold, when there are no unscaled sources.

{\bf Theorem} (scaled sources).  The equation:
\begin{equation}
\triangle_\gamma \varphi - r(\gamma)\varphi = f(.,\varphi) \equiv -a\varphi^{-P} -
q\varphi^{-P^{\prime}}+b\varphi^Q
\end{equation}
with $a\geq 0$, $q\geq 0$, $b\geq 0$; $a, q, b \in W^p_{s,\delta+2}$, $s>\frac{n}{p}$, 
$-\frac{n}{p}<\delta <n-2-\frac{n}{p}$, has a solution $\varphi = 1+u$, $u \in
W^p_{s+2,\delta}$, $\varphi>0$ which can be obtained constructively, if either
(a.) or (b.) holds:

(a.)  On the subset where $r(\gamma)<0$
\begin{equation}
|r(\gamma)|\leq b \; ,
\end{equation}

(b.)  $(M,\gamma)$ is in the positive Yamabe class.  

The solution is unique in either case.

{\it Proof.}

(a.)  The solution exists, with the indicated properties, because the equation
admits a subsolution $\varphi_-$, with $0<\varphi_-$, which is the solution of the
equation (from section 7) 
\begin{equation}
\triangle_\gamma\varphi_- - r(\gamma)\varphi_- -b\varphi^Q_- =0 .
\end{equation}
The solution satisfies $\varphi_- \leq 1$ under the hypothesis made on $r(\gamma)$
because the equation for $\varphi_-$ admits then a supersolution equal to 1.  The
original Lichnerowicz equation (81) admits as supersolution $\varphi_+ \geq 1$ the
solution of the equation (cf. section 8)
\begin{equation}
\triangle_\gamma\varphi_+ + a\varphi^{-P}_+ +q\varphi^{-P^\prime}_+ = 0
\end{equation}
because we have
\begin{equation}
r(\gamma)\varphi_+ +b\varphi^Q_+ \geq 0  \; \mbox{ if } \; \varphi_+\geq 1 \; \mbox{ and }
\; r(\gamma)+b\geq 0.
\end{equation}

(b.)  When $(M,\gamma)$ is in the positive Yamabe class, the equation is equivalent to an
equation of the same type with zero linear term because of conformal covariance.  We may
then argue existence just as in (a.), because the condition on $r(\gamma)$ when it is negative
has become vacuous.

The solution tending to $1$ at infinity of the equation 
with $r(\gamma)=0$ is unique because of the monotonicity of $f$ in 
$\varphi$.  In the general case one uses the conformal transformation of
curvature as in Section \ref{BrillCantor}. Take for simplicity of writing q=0.
We have now if $\varphi_i$, $i=1$ or 2, is a solution,
\begin{equation}
-r(\varphi_i^{2p} \gamma) = -b + a \varphi_i^{-P-Q} \; ;
\end{equation}
therefore the conformal identity with $u=\varphi_1^{-1} \varphi_2$ gives 
\begin{equation}
\triangle_{\varphi_1^{2p}\gamma} u + (b - a\varphi_1^{-(-P+Q)}) u
= (b-a\varphi_2^{-(P+Q)}) u^Q \; .
\end{equation}
This equation may be written
\begin{equation}
\triangle_{\varphi_1^{2p} \gamma} u - \left\{b \left(\frac{u^{Q-1}-1}{u-1}\right)
+ a \varphi_{1}^{-(P+Q)} \left( \frac{1-u^{-P-1}}{u-1}\right)\right\} u(u-1) = 0
\end{equation}
If $u>0$, $b\geq0$, $a\geq0$ the function u, which tends to 1 at infinity,
can only be $u \equiv 1$ on $M$.

{\it Remark 1.} We see that the condition that $(M,\gamma)$ be in the positive Yamabe class
is not necessary for the existence of a positive solution if $b\not\equiv 0$.
However if $b \not\equiv 0$ the Hamiltonian constraint is coupled with the momentum
constraint, and its solution is not the whole story.

{\it Remark 2.}  The condition $b\geq -r(\gamma)$ will somewhat 
be relaxed in the last section but we will require $b>0$.

\section{Discontinuous Sources}
\label{DiscontinuousSources}

It is essential for physical applications to admit isolated sources, hence discontinuous
functions $q$.  This possibility is included if we extend the previous existence theorem
to functions $q\in W^p_{0,\delta +2}$.  We also will take $a\in W^p_{0,\delta +2}$ to 
include the possibility of discontinuous scaled momentum v.  We take $d=b\in 
W^p_{s,\delta +2}$, $s>\frac{n}{p}$.  We leave more general cases for later study.

{\bf Theorem.}  The Lichnerowicz equation with scaled sources,
\begin{equation}
\triangle_\gamma\varphi - r(\gamma)\varphi =f(.,\varphi)\equiv -a\varphi^{-P}-q
\varphi^{-P^\prime}+b\varphi^Q \; ,
\end{equation}
on a $(p,\sigma,\rho)$, $\sigma>\frac{n}{p}+2 \; , \; \rho>-\frac{n}{p}$ asymptotically
euclidean manifold $(M,\gamma)$ in the positive Yamabe class has one and only one
solution $\varphi>0 \; , \; \varphi-1=u\in W^p_{2,\delta}$ if $a,q\in W^p_{0,\delta +2}$\
with $\delta>-\frac{n}{p} \; , \; p>\frac{n}{2} \; , \; a\geq 0 \; , \; q\geq 0 \; , \;
b\geq 0 \; , \; b\in W^p_{s,\delta +2} \; , \; s>\frac{n}{p}$.

{\it Proof.}  We first conformally transform the equation to an equation with no linear
term. We then proceed as follows.  Consider a Cauchy sequence $a_{\nu}, q_{\nu}
\in W^p_{s,\delta+2}, \; s>\frac{n}{p}$, converging in the $W^p_{0,\delta +2}$ norm to
$a, \; q$.  Denote by $\varphi_{\nu}=1+u_{\nu}$ the solution with 
coefficients $a_{\nu}, \; q_{\nu}$.  We know that $u_{\nu}\in
W^p_{s+2,\delta}$ and that there exists numbers $\ell >0$ (depending only on 
$(M,\gamma)$ and $b$) and $m\geq \ell$ (depending only on $(M,\gamma)$ and the 
$W^p_{0,\delta+2}$ norms of $a$ and $q$) such that $\ell \leq \varphi_{\nu}
\leq m$.

The difference $u_{\nu}-u_\mu$ satisfies the equation:
\begin{equation}
\triangle_{\gamma}(u_{\nu}-u_\mu)-A_{\mu{\nu}}(u_{\nu}-u_\mu)=
\varphi^{-P}_\mu(a_{\nu}-a_\mu)+\varphi^{-P^\prime}_\mu(q_{\nu}
-q_\mu)
\end{equation}
with:
\begin{equation}
A_{\mu \nu}\equiv a_\nu \left(\frac{\varphi^{-P}_\mu -\varphi^{-P}_{\nu}}{\varphi_{\nu}-
   \varphi_\mu}\right)+q_{\nu} \left(\frac{\varphi^{-P^\prime}_\mu 
-\varphi^{-P^\prime}_{\nu}}{\varphi_{\nu}-\varphi_\mu} \right)+b \left(
\frac{\varphi^Q_{\nu}-\varphi^Q_\mu}{\varphi_\nu -\varphi_\mu} \right)
\end{equation}
Recall that for $n=3$ we have $P=7$, $P^\prime=3$ and $Q=5$.  The quotients in the
above formulas are then polynomials (with coefficients equal to 1) in $\varphi_\mu^{-1}$ and
$\varphi^{-1}_\nu$ for the first two, and $\varphi_\mu$ and $\varphi_\nu$ for the
third.  Therefore, they are on the one hand positive and, on the other hand, uniformly
bounded (for any pair $\nu, \; \mu$) because $0< \ell \leq \varphi_\mu, \; \varphi_\nu
\leq m$.  For general $n$ the numbers $P, \; P^\prime$ and $Q$ are positive rationals,
the quotients in the formula are also positive and uniformly bounded.  We deduce
from this uniform boundedness that there exists a number $N$ such that 
\begin{equation}
\|A_{\mu\nu} \|_{W^p_{0,\delta + 2}}\leq N \left\{ \|a_\nu \|_{W^p_{0,\delta + 2}}+
\|q_\nu \|_{W^p_{0,\delta + 2}}+\|b\|_{W^p_{0,\delta + 2}}\right\}.
\end{equation}
We infer from this estimate and the positivity of $A_{\mu\nu}$ that the operator
$\triangle_\gamma - A_{\mu\nu}$ is injective in $W^p_{2,\delta}$ 
(See Condition 2 in Theorem 1 of Appendix B).  Therefore,
there exists a number $C$ depending only on
$(M,\gamma)$, the $W^p_{s,\delta +2}$ norm of $b$, and the $W^p_{0,\delta+2}$ norms
of $a$ and $q$ such that 
\begin{equation}
\|u_\nu - u_\mu \|_{W^p_{2,\delta}}\leq C \left\{\|a_\nu - a_\mu \|_{W^p_{0,\delta+2}}
+ \|q_\nu - q_\mu \|_{W^p_{0,\delta+2}}\right\}.
\end{equation}
Since $a_\nu$ and $q_\nu$ are Cauchy sequences, the same is true of $u_\nu$, because of 
the above inequality.  Hence $u_\nu$ converges in $W^p_{2,\delta}$ to a limit 
$u\in W^p_{2,\delta}$.  The convergence is {\it a fortiori} in $C^0_\alpha$
if $p>\frac{n}{2}$, and for some positive $\alpha$, since $\delta >-\frac{n}{p}$.
The $u_\nu$'s are such that $1+u_\nu \geq \ell>0$; therefore also $\varphi=1+u \geq
\ell >0$.  The function $\varphi$ satisfies the Lichnerowicz equation (in the sense
of generalized derivatives) with scaled sources.  

\section{General Cases}
\label{GeneralCases}

In the case where there are unscaled sources the coefficient $d$ in the Lichnerowicz
equation is negative or zero on a maximal initial manifold $M$.  It can take different
signs if $M$ is not maximal.  The previous simple method to obtain sub and super
solutions does not apply.  We will then look for constant sub and super solutions 
$\ell$ and $m$, $0<\ell \leq 1 \leq m$.  We will also obtain a new theorem 
for the Lichnerowicz equation in the case
of scaled sources on a non maximal submanifold.  To make the algebra easier we restrict
our study to the important physical case $n=3$.  
Results along the same lines can likely be obtained for general n. The equation is then
\begin{eqnarray}
\triangle_\gamma \varphi - r\varphi +a\varphi^{-7} +q\varphi^{-3} - d\varphi^5 =0, \nonumber\\
a\geq 0, \; q\geq 0, \; d=b-c, \; b\geq 0, \; c\geq 0 .
\end{eqnarray}
The numbers $\ell$ and $m$ are admissible sub and supersolutions if they satisfy on $M$
the following inequalities:
\begin{equation}
P_x(\ell^4)\leq 0, \; \; P_x(m^4)\geq 0, \; \; \mbox{for all } x\in M, \; 0<\ell \leq 1 \leq m
\end{equation}
where $P_x$ is the polynomial
\begin{equation}
P_x(z)\equiv d(x)z^3 + r(x)z^2 - q(x)z - a(x).
\end{equation}

{\it Remark.}  In the case of $n>3$ the problem is the study of the sign of the function:
\begin{equation}
F_x(z)\equiv d(x)z^n + r(x)z^{n-1} - q(x)z^{(n-1)/2} - a(x)
\end{equation} 
for numbers $\ell^{4/(n-2)}$ and $m^{4/(n-2)}$.  

Since all the coefficients in $P_x$ tend to zero at infinity the conditions that we will
obtain depend on the ratios of their respective decays.

We denote by $M_+$ the subset of $M$ where $d>0$, by $M_-$ the subset where $d<0$, by
$M_0$ the subset where $d=0$.  In the case of isolated sources $M_-$ is a compact subset
of $M$.  We study the sign of $P_x$ on these various subsets.  The derivative of $P_x$ is
\begin{equation}
dP_x/dz = 3d(x)z^2 + 2r(x)z - q(x).
\end{equation}

1.  On $M_+$, $d(x)>0$, the derivative $dP_x/dz$ has 2 roots of opposite signs.  The 
positive root is 
\begin{equation}
\zeta_+ (x) = \frac{-r(x)+(r^2(x)+3d(x)q(x))^{1/2}}{3d(x)}\geq 0.
\end{equation}
We have $\zeta_+(x)>0$ if $r(x)<0$, or if $r(x)\geq 0$ and $q(x)>0$.

$dP_x/dz$ is equal to $-q(x)\leq 0$ for $z=0$ and is negative or zero as long as 
$z\leq \zeta_+(x)$.  Therefore $P_x$ decreases from $a(x)\leq 0$ for $z=0$ to a minimum
for $z=\zeta_+(x)$ and then increases to $+\infty$ when $z$ increases to $+\infty$.  Hence
$P_x$ has one and only one positive root $z_+(x)$.  We have $P_x(z)\geq 0$ as long as 
$z\geq z_+(x)$.

There exists $\ell(x)>0$ such that $P_x(\ell^4(x))\leq 0$ if and only if $z_+(x)>0$.  Indeed
numbers $\ell(x)$ and $m(x)$ such that
\begin{equation}
0<\ell(x)\leq z_+(x)\leq m(x) \; , \; x\in M_+
\end{equation}
satisfy
\begin{equation}
P_x(\ell^4(x))\leq 0, \; \; P_x(m^4(x))\geq 0.
\end{equation}
{\bf Lemma 1.}  There exist numbers $\ell_+$ and $m_+$ such that: 
\begin{equation}
P_x(\ell^4_+)\leq 0, \; \; P_x(m^4_+)\geq 0 \mbox{ for all } x\in M_+
\end{equation}
if and only if
\begin{equation}
\inf_{x\in M_+} z_+(x)>0.
\end{equation}
and
\begin{equation}
\sup_{x\in M_+} z_+(x)<+\infty.
\end{equation}
Sufficient conditions for the first inequality are
\begin{equation}
\inf_{x\in M_+}\left\{\frac{-r(x)}{3d(x)} + \left(\frac{r^2(x)}{9d^2(x)} +
\frac{q(x)}{3d(x)}\right)^{1/2}\right\}>0
\end{equation}
or
\begin{equation}
\inf_{x\in M_+} \frac{a(x)}{d(x)+|r(x)|}>0.
\end{equation}
For the second inequality they are that $\frac{|r(x)|}{d(x)}, \; \frac{q(x)}{d(x)}, \;
\frac{a(x)}{d(x)}$ be uniformly bounded on $M_+$.

{\it Proof.}  The necessary condition as well as the first sufficient condition are 
consequences of the previous study.  Sufficient conditions for this first condition
to be satisfied are that one of the two terms in the sum has a strictly positive
infimum.  The second sufficient condition results from the fact (elementary calculus)
that $P_x(z)\leq 0$ if 
\begin{equation}
z\leq \min \left(1,\frac{a(x)}{d(x)+|r(x)|}\right)
\end{equation}

{\it Remark.}  The sufficient conditions will be satisfied on the whole of $M_+$ if
we can split it into two subsets, $M_+ \equiv M_1 \cup M_2$, such that
\begin{equation}
\inf_{x\in M_1} \frac{a(x)+q(x)}{d(x)+|r(x)|}>0 \mbox{  and } \inf_{x\in M_2}
\frac{-r(x)}{d(x)}>0.
\end{equation}
This pair of inequalities can be realized when $M$ is compact and $a(x) + q(x)\not\equiv 0$
by a conformal change of choice of the metric $\gamma$ to a metric $\gamma^\prime$ 
having a strictly negative curvature in the complement of $M_1$ in $M$.  Such a 
construction can also eventually be made in the asymptotically flat case, by resolution
of an adequate Dirichlet problem.  

2.  On $M_-$, $d(x)<0$.

We have $P_x(z)<0$ for all $z>0$, hence no admissible $m(x)$, if $r(x)\leq 0$.  We 
therefore suppose $r(x)>0$ for all $x \in M_-$.  If $r^2(x)+3q(x)d(x)\leq 0$, we have
$dP_x/dz\leq 0$ for all $z$; and the polynomial $P_x$ takes non negative values only
if it is identically zero.  If $r^2(x)+3q(x)d(x)>0$ the polynomial $dP/dz$ has two
positive roots:
\begin{eqnarray}
0\leq \zeta_1(x)&=& \frac{r(x)-\left\{r^2(x)-3q(x)|d(x)|\right\}^{1/2}}{3|d(x)|}, \\
\zeta_2(x)&=& \frac{r(x)+\left\{r^2(x)-3q(x)|d(x)|\right\}^{1/2}}{3|d(x)|}>0,
\end{eqnarray} 
with $\zeta_1(x)>0$ if and only if $q(x)\neq 0$.

The polynomial $P_x$ decreases for $0\leq z \leq \zeta_1(x)$, increases for 
$\zeta_1(x)\leq z \leq \zeta_2(x)$, and decreases to $-\infty$ for $z\geq \zeta_2(x)$.  We have
$P_x(0)=-a(x)\leq 0$.  Therefore $P_x$ takes negative values for some $z>0$ if either 
$a(x)>0$ or $\zeta_1(x)>0$, {\it i.e.} $q(x)>0$.  The polynomial $P_x$ takes positive
values, equivalently admits two positive roots $z_1(x)$ and $z_2(x)$ which are such that
\begin{equation}
\zeta_1(x)\leq z_1(x)\leq \zeta_2(x)\leq z_2(x),
\end{equation}
if and only if its maximum, attained for $z=\zeta_2(x)$, is positive,
\begin{equation}
P_x(\zeta_2(x))\geq 0.
\end{equation}
We have then $P_x(z)\leq 0$ for $0\leq z \leq z_1(x)$, and $P_x(z)\geq 0$ for 
$z\geq z_2(x)$.  If
\begin{equation}
r^2(x)+3q(x)d(x)\leq 0 \; ,
\end{equation} 
the polynomial $P_x$ is always decreasing.  It takes positive ({\it i.e.}, non-negative)
values only if it is identically zero.

{\bf Lemma 2.}  Suppose that $r(x)>0$, $r^2(x)-3q(x)d(x)>0$ and 
$P_x(\zeta_2(x))\geq 0$
for all $x\in M_-$.  There exist numbers $\ell_-$ and $m_-$ such that:
\begin{equation}
P_x(\ell^4_-)\leq 0, \; \; P_x(m^4_-)\geq 0 \quad \mbox{for all } x\in M_-
\end{equation}
if the following conditions are satisfied:
\begin{equation}
\inf_{M_-}z_1(x)>0, \; \; \sup_{x\in M_-}z_1(x)\leq \inf_{x\in M_-}z_2(x),
\end{equation}
and $\frac{|r(x)|}{|d(x)|}, \; \frac{q(x)}{|d(x)|}, \; \frac{a(x)}{|d(x)|}$ are
uniformly bounded on $M_-$.

{\it Proof.}  All numbers $\ell_-$ and $m_-$ such that:
\begin{equation}
\ell_-\leq z_1(x), \quad z_1(x)\leq m_-\leq z_2(x) \mbox{  for all } x\in M_-
\end{equation}
are such that $P_x(\ell^4_-)\leq 0, \; P_x(m^4_-)\geq 0$.  These numbers exist, with
$\ell_->0$ and $+\infty\geq m_-\geq \ell_-$ under the given conditions.  

3.  On $M_0, \; d(x)=0, \; P_x$ reduces to a second order polynomial
\begin{equation}
P_x(z)=r(x)z^2-q(x)z-a(x).
\end{equation}
If $r(x)\leq 0$, then $P_x<0$ as soon as $z>0$ except if it is identically zero.  We
suppose $r(x)>0$.  Then $P_x$ admits one positive root $z_0(x)$:
\begin{equation}
z_0(x)=(2r(x))^{-1}\left\{q^2(x)+4a(x)r(x)\right\}^{1/2}\geq 0.
\end{equation}
{\bf Lemma 3.}  We suppose that $r(x)>0$ for all $x\in M_0$.  

There exist $\ell_0>0$ and $m_0\geq \ell_0$ such that $P_x(\ell^4_0)\leq 0$ and 
$P_x(m^4_0)\geq 0$ for all $x\in M_0$ if and only if
\begin{equation}
\inf_{x\in M_0}\frac{a(x)}{r(x)}>0 \quad \mbox{or} \quad  \inf_{x\in M_0}\frac{q(x)}{r(x)}>0
\end{equation}
and
\begin{equation}
\sup_{x\in M_0} \frac{a(x)}{r(x)}<+\infty \quad \mbox{and} \quad \sup_{x\in M_0} 
\frac{q(x)}{r(x)}<+\infty.
\end{equation} 
{\it Proof.}  Under one or the other of the first inequalities we have
\begin{equation}
\inf_{x\in M_0} z_0(x)>0.
\end{equation}
The other ones insure 
\begin{equation}
\sup_{x\in M_0} z_0(x)<+\infty.
\end{equation}
We set
\begin{equation}
\ell^4_0=\inf_{M_0}z_0(x), \quad m^4_0=\sup_{M_0}z_0(x).
\end{equation}
All numbers $\ell$ and $m$ satisfying the following inequalities
\begin{equation}
0<\ell\leq \ell_0 = \inf_{M_0}z_0(x)\leq \sup_{x\in M_0}z_0(x) = m_0\leq m
\end{equation}
satisfy $P_x(\ell^4)\leq 0$ and $P_x(m^4)\geq 0$ for all $x\in M_0$.

The following lemma is an immediate consequence of the previous three.

{\bf Lemma 4.}  We suppose that the conditions given in the lemmas 1, 2,
and 3 for the existence of $\ell_- , \; \ell_+ , \; \ell_0$ and $m_- , \;
m_+ , \; m_0$ are satisfied. Then there exists $\ell$ and $m$ such that:
\begin{equation}
0<\ell \leq 1 \leq m \mbox{ and } P_x(\ell^4)\leq 0 \; , \quad 
P_x(m^4)\geq 0 \quad \mbox{ for all } x\in M
\end{equation}  
if the following inequalities hold:
\begin{equation}
m_+ \leq m_- \; , \quad m_0 \leq m_- \; , \quad m_- \geq 1 \; .
\end{equation}

{\it Proof.}  Take
\begin{equation}
m=m_- \; , \quad \ell =\min(1,\ell_0,\ell_+,\ell_-).
\end{equation}
Then $\ell$ and $m$ satisfy the required inequalities for all $x\in M$.
They are admissible sub and supersolutions.

{\bf Theorem.}  On a 3-dimensional asymptotically euclidean manifold the
Lichnerowicz equation
\begin{eqnarray}
\triangle_\gamma\varphi - r\varphi + &a\varphi^{-7}& + q\varphi^{-3} - d
\varphi^5 = 0 \; , \nonumber \\
a\geq 0 \; , \quad q\geq 0 \; , \quad d&=&b-c \; , \quad b\geq 0 \; , \quad
c\geq 0 \; .
\end{eqnarray}
with $r, \; a, \; q, \; d\in W^p_{s,\delta+2}, \; s>\frac{n}{p}, \; -\frac{n}{p}
<\delta<n-2-\frac{n}{p}$, admits a solution $\varphi>0, \; \varphi - 1\in
W^p_{s+2,\delta}$ if the assumptions of Lemma 4 are satisfied.

{\bf Corollary}  {\it No Unscaled Sources, $d\equiv b>0$.}  The Lichnerowicz equation has
a solution $\varphi >0$, $\varphi - 1\in W^p_{s+2,\delta}$ if

(i)  The quotients $\frac{|r(x)|}{b(x)}, \; \frac{q(x)}{b(x)}, \; \frac{a(x)}{b(x)}$
are uniformly bounded.

(ii)  There is a positive number $\epsilon >0$ such that if 
$\frac{a(x)+q(x)}{b(x)}<\epsilon$, then
\begin{equation}
r(x)<0 \mbox{ and } \frac{|r(x)|}{b(x)}>\epsilon^\prime >0 .
\end{equation}
This last condition can be achieved if $a+q\not\equiv 0$ by a 
conformal transformation and solution of a Dirichlet problem in the subset of $M$
where $(a+q)/d<\epsilon$, so long as this subset is compact
(cf. \cite{cbyo1980,I1995}). 

\section{Unscaled Sources, Case $n=3$}

We treat in this section the hamiltonian constraint for unscaled sources in the
case $n=3$.  The Lichnerowicz equation reads
\begin{equation}
\triangle_\gamma\varphi - r\varphi + a\varphi^{-7} + c\varphi^5 = 0
\end{equation}
The functions $a\geq 0$ and $c\geq 0$ are given on $(M,\gamma)$.

{\bf Theorem.}  Let $(M,\gamma)$ be a $(p,\sigma,\rho)$ asymptotically euclidean
manifold, $\sigma > \frac{n}{p}+2, \; \sigma > -\frac{n}{p}$ with $r>0$.  Let
$a, \; c \in W^p_{s,\delta+2}$ be given on $(M,\gamma)$, $s>\frac{n}{p}, \;
\sigma>-\frac{n}{p}$.  There exists an open set of values of $a$ and $c$ such that
the Lichnerowicz equation with unscaled sources has a solution $\varphi >0$, with
$1-\varphi \in W^p_{s+2,\delta}$.

{\it Proof,}  We look for constant admissible sub and supersolutions $\ell$ and $m$
such that
\begin{eqnarray}
&0<\ell\leq 1\leq m \; ,& \nonumber\\
P_x(\ell^4)\geq 0, \quad &P_x(m^4)\leq 0,& \quad \mbox{for all } x\in M,
\end{eqnarray}
where $P$ is the polynomial,
\begin{equation}
P_x(z)\equiv c(x)z^3 - r(x)z^2 + a(x).
\end{equation}

1.  Case $c>0$.

We set $z=X^{-1}$ and consider the polynomial which  has the same sign as $P_x$,
\begin{equation}
Q(X)\equiv a\left\{X^3 - a^{-1}rX + a^{-1}c\right\}.
\end{equation}
This polynomial has 3 real roots if
\begin{equation}
4r^3\geq 27ac^2 \; .
\end{equation}
Two of these roots are positive, given by the classical formulas:
\begin{equation}
X_2\equiv \lambda\sin\frac{\theta}{3} \; , \quad X_1= \lambda\sin\frac{\theta + 2\pi}{3}
\end{equation}
with:
\begin{equation}
\lambda = \frac{2r^{1/2}}{(3a)^{1/2}} \; , \quad \sin\theta = \frac{3c(3a)^{1/2}}{2rr^{1/2}}
\; , \quad 0\leq \theta \leq \frac{\pi}{2} \; .
\end{equation}
The corresponding roots of $P_x$ are
\begin{equation}
z_1\equiv \frac{1}{X_1}\leq \frac{1}{X_2}\equiv z_2(x).
\end{equation}
We have $P_x(z)\geq 0$ for $0\leq z\leq z_1(x), \; \; P_x(z)\leq 0$ for $z_1(x)\leq z\leq z_2(x)$.

2.  Case $c(x)=0$.

The polynomial $P_x$ reduces to:
\begin{equation}
P_x(z)\equiv -r(x)z^2 + a(x).
\end{equation}
We have $P_x(z)\geq 0$ for $0\leq z\leq(r^{-1}a)^{1/2}$, and $P_x(z)\leq 0$ for 
$z\geq (r^{-1}a)^{1/2}$.  Note that $(r^{-1}a)^{1/2}$ is the value for $c=0$ of the 
previously computed $z_1$ while the previous $z_2$ tends to infinity when $c$ tends
to zero.  The cases $c(x)\geq 0$ are thus unified.

The following constants $\ell$ and $m$ are sub and supersolutions if
\begin{equation}
\ell\leq z_1(x), \quad z_1(x)\leq m \leq z_2(x) \quad \mbox{for all } x\in M.
\end{equation}
They exist, satisfying the required properties $0<\ell \leq 1\leq m$, if
\begin{equation}
\inf_{x\in M}z_1(x)>0 \; , \quad \inf_{x\in M}z_2(x)\geq 
\max\left\{1,\sup_{x\in M}z_1(x)\right\}.
\end{equation}
One then takes
\begin{equation}
\ell = \min\left\{1,\inf_{x\in M}z_1(x)\right\}, \quad m=\inf_{x\in M}z_2(x).
\end{equation}
We give below sufficient conditions to satisfy the various inequalities, using
the expressions
\begin{equation}
z_1 = \frac{2}{\sqrt{3}}\frac{r^{-1}a}{\sin((\theta + 2\pi)/3)}, \quad
z_2 = \frac{2}{\sqrt{3}}\frac{r^{-1}a}{\sin(\theta /3)},\quad \mbox{with } 
0\leq \theta \leq \frac{\pi}{2}.
\end{equation}
The functions $\sin(\theta/3)$ and $\sin((\theta + 2\pi)/3)$ are respectively increasing
and decreasing when $\theta$ increases from $0$ to $\frac{\pi}{2}$.  Denote by $\theta_{\min}$
and $\theta_{\max}$ the infimum and supremum of $\theta$ on $M$ defined as solutions 
between $0$ and $\frac{\pi}{2}$ of the equations
\begin{equation}
\sin\theta_{\min} = \inf_M \frac{3c(3a)^{1/2}}{2rr^{1/2}}, \quad \sin\theta_{\max} = 
\sup_M \frac{3c(3a)^{1/2}}{2rr^{1/2}}.
\end{equation}
Therefore:
\begin{eqnarray}
\inf_M z_1 &\geq& \frac{2}{\sqrt{3}}\inf_M 
(r^{-1}a)\left\{\sin((\theta_{\min} + 2\pi)/3)\right\}^{-1} \nonumber\\
\sup_M z_1 &\leq& \frac{2}{\sqrt{3}}\sup_M
(r^{-1}a)\left\{\sin((\theta_{\max} + 2\pi)/3)\right\}^{-1}  \nonumber\\
\inf_M z_2 &\geq& \frac{2}{\sqrt{3}}\inf_M
(r^{-1}a)\left\{\sin\theta_{\max}\right\}^{-1}.
\end{eqnarray}
We find by elementary calculus that:
\begin{equation}
0\leq \sin((\theta_{\max} + 2\pi)/3) - \sin(\theta_{\max}/3) = 
\sqrt{3}\cos((\theta_{\max} + \pi)/3)\leq
\frac{\sqrt{3}}{2}
\end{equation}
The minimum zero is attained for $\theta_{\max} = \frac{\pi}{2}$, {\it i.e.}
\begin{equation}
\sup_M \frac{3c(3a)^{1/2}}{2rr^{1/2}} = 1 \; .
\end{equation}
To insure the existence of constants $\ell$ and $m$ satisfying the required
properties we suppose 
\begin{equation}
\ell_- = \frac{2}{\sqrt{3}}\inf_M (r^{-1}a)\left\{\sin((\theta_{\min} + 2\pi)/3)
\right\}^{-1}>0 \; ,
\end{equation}
and we set 
\begin{equation}
\ell = \min\{\ell_-,1\}.
\end{equation}
We suppose also
\begin{equation}
m_+\equiv \frac{2}{\sqrt{3}}\inf_M(r^{-1}a)\left\{\sin(\theta_{\max}/3)\right\}^{-1}\geq 1
\end{equation}
and
\begin{equation}
m_-\equiv \frac{2}{\sqrt{3}}\sup_M(r^{-1}a)\left\{\sin((\theta_{\max} +2\pi)/3)\right\}^{-1}
\leq m_+ \; .
\end{equation}
The condition $m_-\leq m_+$ can be satisfied for an open set of values of the coefficients
$c$, $a$ (given $r$) due to the previous elementary study.  We can take for $m$ any number between
$\max\{1,m_-\}$ and $m_+$.  The numbers $\ell$ and $m$ so chosen are admissible sub and 
supersolutions of the Lichnerowicz equation.  The existence of a solution $\varphi$ with the
required properties results from the general theorem, given in Section 5.

\section{Coupled System}

In the conformal method the momentum and the hamiltonian constraints decouple when the
initial manifold $M$ has constant mean extrinsic curvature 
and the unscaled sources have a momentum $N=0$.  The theorems of the previous sections
are then sufficient to give existence, non-existence or uniqueness theorems of the
systems of constraints.  We will in the next sections study cases where one of these
hypothesis does not hold; hence the constraints do not decouple.

\section{Implicit Function Theorem Method}

The use of the implicit function theorem is the simplest way of proving existence of solutions
of equations in the neighbourhood of a given one.  It works as follows.

Let $U$ and $V$ be open sets of Banach spaces $X$ and $Y$ and let ${\cal F}$ be a $C^1$
mapping from $U\times V$ into another Banach space $Z$:
\begin{equation}
{\cal F}:U\times V\rightarrow Z \mbox{ by }(x,y)\mapsto {\cal F}(x,y) \; .
\end{equation}
Suppose that the partial derivative of ${\cal F}$ with respect to $y$ at a point
$(x_0,y_0)\in U\times V, \;  {\cal F}^\prime_y(x_0,y_0)$, is an isomorphism from
$Y$ onto $Z$; then there exists a neighbourhood $W$ of $x_0$ in $U$ such that the equation
\begin{equation}
{\cal F}(x,y) = 0
\end{equation}
has a solution $y\in V$ for each $x\in W$.

We consider the quantities $q$ and $v$ (scaled sources) together with $N$ and $u$, a
traceless symmetric 2-tensor as given on the asymptotically euclidean manifold $(M,\gamma)$,
with $q, \; v, \; 1-N^{-1}\in W^p_{s,\delta + 2}, \; u\in W^p_{s+1,\delta + 1}$.  We will
discuss the existence of $\varphi$ and $\beta$ as we perturb $J$ and $\tau$
away from zero.  The
points $x,y$ and the Banach spaces $X,Y,$ and $Z$ are as follows:
\begin{eqnarray}
x &\equiv& (\tau,J)\in X \equiv W^p_{s+1,\delta +1}\times W^p_{s,\delta +2} \; , \nonumber\\
y &\equiv& (\beta ,\varphi -1)\in V \equiv Y\cap \{\varphi >0\}, \; Y\equiv
W^p_{s+2,\delta}\times W^p_{s+2,\delta} \; . \nonumber\\
Z &\equiv& W^p_{s,\delta +2} \times W^p_{s,\delta +2} \; .
\end{eqnarray}
The mapping ${\cal F}$ is given by 
\begin{equation}
{\cal F}(x,y)\equiv ({\cal H}(\tau,J;\varphi ,\beta), \; {\cal M}(\tau,J;\varphi ,\beta))
\end{equation}
where ${\cal H}$ and ${\cal M}$ are the left hand sides of the conformal formulation of
the constraints, ${\cal H}(x,y)\equiv \triangle_\gamma\varphi - f(.;\tau;\beta,\varphi), \quad
{\cal M}(x,y)\equiv \bar{\nabla}.((2N)^{-1}{\cal L}\beta) - h(.;\tau,J;\varphi)$.

The multiplication properties of weighted Sobolev spaces show that $\cal F$ is a $C^1$ mapping
from $X\times V$ into $Z$ if $s>\frac{n}{p}$ and $\delta > -\frac{n}{p}$.  The partial derivative
${\cal F}^\prime_y$ at a point $(x,y)$ is the linear mapping from $Y$ into $Z$ given by:
\begin{equation}
(\delta \beta,\delta \varphi)\mapsto ({\cal H}^\prime_y, {\cal M}^\prime_y).(\delta \beta, 
\delta \varphi)
\end{equation}
with ($A$ is given by (\ref{Aij}))
\begin{eqnarray}
{\cal H}^\prime_y.(\delta \beta,\delta \varphi)&\equiv& \triangle_\gamma \delta \varphi - 
\alpha \delta\varphi + 2k_n\varphi^{-P} 2 N^{-1} A.
{\cal L} \delta \beta \; , \nonumber\\
\alpha &\equiv& r + P\varphi^{-P-1}a(\beta) +P^\prime q\varphi^{-P^\prime -1}+dQ\varphi^{Q-1} \; ,
\nonumber\\ \nonumber\\
{\cal M}^\prime_y.(\delta \beta,\delta\varphi) &\equiv& 
\nabla.(2N^{-1} \cal{L} \delta\beta) -\lambda \delta \varphi \; , \nonumber \\
\lambda &\equiv& \frac{2(n-1)}{(n-2)}\varphi^{(n+2)/(n-2)}D\tau + \frac{2(n+2)}{n-2}
\varphi^{(n+6)/(n-2)}J \; .\nonumber\\
\end{eqnarray}

{\bf Theorem.}  Specify on the asymptotically euclidean manifold $(M,\gamma)$ a
traceless tensor $u\in W^p_{s+1,\delta +1}$, a scalar $N>0$ with $N-1\in W^p_{s+2,\delta}$,
and scaled and unscaled sources $q, \; v, \; c\in W^p_{s,\delta +2}, \; s>\frac{n}{p}, \;
-\frac{n}{p}<\delta <n-2-\frac{n}{p}$.  Let $(\beta_0,\varphi_0)$ be a solution of the 
corresponding constraints with $D\tau_0 = 0$ (hence $\tau_0 = 0$ since $\tau_0\in
W^p_{s+1,\delta +1}$), and $J_0=0$.  Suppose that on $M$
\begin{equation}
\alpha_0 \equiv r+P\varphi^{-P-1}_0 a(\beta_0) + P^\prime\varphi^{-P^\prime -1}_0 q
-cQ\varphi^{Q-1}_0\geq 0 \; .
\end{equation}
Then there exists a neighbourhood $U$ of $(\tau_0,J_0)$ in $X$ such that the coupled
constraints have one and only one solution $(\beta,\varphi), \; \varphi>0, \; 
(\beta,1-\varphi)\in Y$. 

{\it Proof}  Under the hypotheses that we have made the partial derivative 
${\cal F}^\prime_y(x_0,y_0)$ is an isomorphism from $Y$ onto $Z$ because the system of 
linear elliptic equations
\begin{eqnarray}
&\bar{\nabla}.\{(2N)^{-1}{\cal L}\delta \beta\}=h& \nonumber \\
&\triangle_\gamma \delta\varphi - \alpha_0\delta \varphi = 
-2k_n\varphi^{-P}(2N^{-1}A).{\cal L}\delta \beta +k&
\end{eqnarray}
has one and only one solution $(\delta \varphi, \delta \beta)\in Y$ for any pair
$(h,k)\in Z$.

{\bf Corollary.}  The conclusion of the theorem holds if $(M,\gamma)$ is in the positive
Yamabe class and $d\geq 0$ (realized in particular if all sources are scaled) without
having to consider the sign of $\alpha_0$.

{\it Proof.}  If $(M,\gamma)$ is in the positive Yamabe class we can choose $(M,\gamma^\prime)$
in the same conformal class and such that $r(\gamma^\prime)=0$.  To the data $N, \; u, \; q, \;
v$ correspond data $N^\prime, \; u^\prime, \; q^\prime, \; v^\prime$ and to the solution
$\beta_0,\varphi_0$ corresponds a solution of the transformed conformal constraints.  The
corresponding $\alpha^\prime_0$ is positive and the conclusion of the theorem applies to the
transformed system, hence also to the original system.  

\section{Constructive Method with Scaled Sources}

We will give in the next two sections another method to obtain solutions of the coupled system.
It will give new results for unscaled sources on a maximal manifold. It is possible, 
though not proven, that the hypotheses we make in the case of scaled sources on a 
non-maximal manifold imply that this manifold is in the positive Yamabe class.

{\bf Lemma 1.}  The equation
\begin{equation}
\triangle_\gamma\varphi = f(.,\varphi)\equiv r\varphi - a\varphi^{-P} - q\varphi^{-P^\prime}
+ b\varphi^Q
\end{equation}
with $r, \; a, \; q, \; b$ satisfying the hypothesis of the theorem in Section 9 admits as a
supersolution the solution $\Phi(A), \; 1-\Phi(A)\in W^p_{s+2,\delta},$ of the equation
\begin{equation}
\triangle_\gamma\varphi = f_A(.,\varphi)\equiv -A\varphi^{-P} - q\varphi^{-P^\prime}
\end{equation}
if $a\leq A$, with $A$ a given function in $W^p_{s,\delta +2}$.

{\it Proof.}  The function $\Phi(A)$ exists by the theorem in Section 9.  It satisfies
\begin{equation}
\triangle_\gamma \Phi - f(.,\Phi) = f_A(.,\Phi) - f(.,\Phi) \leq (a-A)\Phi^{-P}\leq 0 \; ,
\end{equation}
hence it is a supersolution.

{\bf Theorem.}  Under the conditions on $r$ and $b$ given in the theorem of Section 9 there
exists a number $\epsilon>0$ such that if 
\begin{equation}
\|D\tau \|_{W^p_{0,\delta +2}}\leq \epsilon \; , \quad n>p, \; \delta >-n/p,
\end{equation}
the coupled constraints admit a solution $(\beta,\varphi)$ with $\beta, \; 1-\varphi \in
W^p_{s+2,\delta}$.

{\it Proof.}  We will construct a sequence $(\varphi_\nu , \beta_\nu)$ by the inductive algorithm
\begin{eqnarray}
\triangle_\gamma\varphi_\nu = f(.,\beta_{\nu-1},\varphi_\nu)&\equiv& r\varphi_\nu - a(\beta_{\nu-1})
\varphi^{-P}_\nu - q\varphi^{-P^\prime}_\nu + b\varphi^Q_\nu \; , \nonumber\\ \nonumber\\
\nabla_i\left\{(2N^{-1})({\cal L}\beta_\nu)^{ij}\right\} = h^j(.,\varphi_\nu) &\equiv& \nabla_i
\left\{(2N)^{-1}u^{ij}\right\}\nonumber\\
&+& \frac{n-1}{n}\varphi^{2n/(n-2)}_\nu \nabla^j\tau + v^j
\end{eqnarray}
with
\begin{equation}
a(\beta)\equiv k_n(2N)^{-2}|-u+{\cal L}\beta |^2 \; , \quad k_n\equiv \frac{n-2}{4(n-1)} \; .
\end{equation}

The equations for the $\varphi_\nu$'s admit all the same subsolution $\varphi_-$, which depends
only on $r$ and $b$.  They admit the same supersolution $\Phi(A)$ if there exists
$A\in W^p_{s,\delta+2}$ such that $a(\beta_{\nu-1})\leq A$ for all $\nu$.

We start for instance from $\beta_0 =0$ and choose $A$ such that
\begin{equation}
A>a(0)\equiv k_n(2N)^{-2}|u|^2 \; .
\end{equation}
Suppose that $\beta_{\nu-1}\in W^p_{s,\delta+2}$ and that $a(\beta_{\nu-1})<A$.  Then $\varphi_\nu$
exists, $1-\varphi_\nu\in W^p_{s+2,\delta}$, and $\varphi_-\leq \varphi_\nu \leq \Phi(A)$.  Also, 
$\beta_\nu \in W^p_{s+2,\delta}$ exists, and there is a constant $C$ (cf. Appendix A)
depending only on 
$(M,\gamma)$ such that
\begin{equation}
\|\beta_\nu\|_{W^p_{2,\delta}}\leq C\left\{\frac{n}{n-1}\|D\tau\|_{W^p_{0,\delta+2}} 
\sup_M\Phi(A)^{2n/(n-2)} + \|P\|_{W^p_{0,\delta+2}}\right\} \; ,
\end{equation}
where $P$ is the given vector
\begin{equation}
P\equiv \nabla.\left\{(2N)^{-1}u\right\} +v \; .
\end{equation}
The weighted Sobolev multiplication theorem and the expression for $a(\beta)$ imply that if
$\beta_\nu \in W^p_{2,\delta}, \; \frac{n}{p}<1, \; \delta>-\frac{n}{p}$, then
\begin{equation}
a(\beta_\nu)\in W^p_{1,\delta^\prime+2} \quad \mbox{for all }\delta^\prime \; \mbox{such that }
\delta^\prime<\delta+(\delta+\frac{n}{p}) \; ,
\end{equation}
and there exists a number $C$ such that
\begin{equation}
\|a(\beta_\nu ) \|_{W^p_{1,\delta^\prime+2}}\leq C\left\{\|u\| ^2_{W^p_{1,\delta+1}} + 
\|\beta_\nu \|^2_{W^p_{2,\delta}}\right\}.
\end{equation}
By the weighted Sobolev inclusion theorem, there exists then another constant $C$ such that
\begin{equation}
\|a(\beta_\nu)\|_{C^0_{\delta^{\prime\prime}}}\leq C\|a(\beta_\nu)\|_{W^p_{1,\delta^\prime +2}}
\end{equation}
for all $\delta^{\prime\prime} < \delta^\prime +2+\frac{n}{p}$, hence also for all 
$\delta^{\prime\prime} < 2\delta +2+\frac{n}{p}$ .

Since $\delta+\frac{n}{p}>0$ there exists a number $\alpha$ such that
\begin{equation}
\delta + 2 + \frac{n}{p}< \alpha <2\delta + 2 + \frac{n}{p} \; .
\end{equation}
We choose for $A$ a function of the form, with $\mu$ some positive constant,
\begin{equation}
A=\mu / \sigma^\alpha \; ,
\end{equation}
where $\sigma \equiv 1+d^2$ (see section 3).
We have $A\in W^p_{0,\delta+2}$  since $\alpha >\delta +2+\frac{n}{p}$ .  For such a function $A$,
the inequality $a(\beta_\nu)\leq A$ is equivalent to
\begin{equation}
\sigma^\alpha a(\beta_\nu)\leq \mu \; , \quad \mbox{\it i.e.}, \; \|a(\beta_\nu)\|_{W_0^2,\alpha}
\leq \mu .
\end{equation}
Using the previous estimates we see that a sufficient condition to insure on $M$ the
inequality $a(\beta_\nu)\leq A$ is to have some number depending only on $(M,\gamma)$ and $N$,
denoted by $C$, that has the property
\begin{equation}
\|D\tau \|^2_{W^p_{0,\delta+2}} \sup_M \Phi^{4n/(n-2)}_\mu + \|v\|^2_{W^p_{0,\delta+2}} + 
\|u\|^2_{W^p_{1,\delta+1}}
\leq C\mu \; ,
\end{equation}
where we have set $\Phi_\mu \equiv \Phi(\sigma^{-\alpha}\mu)$.

We choose $\mu$ large enough to have:
\begin{equation}
C\mu >S \; , \quad S\equiv \|v\|^2_{W^p_{0,\delta+2}} + \|u\|^2_{W^p_{0+1,\delta+1}} \; .
\end{equation}
The inequality obtained above shows that we can construct $\varphi_{\nu+1}$ and hence
$\beta_{\nu+1}$, enjoying the same properties as $\varphi_\nu , \; \beta_\nu$ if $D\tau$
is sufficiently small in $W^p_{0,\delta+2}$ norm.  The existence of a solution $(\varphi,\beta)$
of the coupled constraints as limit of a subsequence is proved by a compactness argument and
elliptic regularity as in the case of the hamiltonian constraint with a given $\beta$.

{\it Remark.}  The number $\epsilon$ depends on the choice of $\mu$ and the function $\Phi_\mu$.
Neither of those depends on $r$ or on $b$, {\it i.e.}, on $\tau$.  However, our theorem imposes 
a restriction on the size of $\tau$ on the subset of the manifold $M$ where $r<0$, since it 
supposes that $b\equiv \frac{n-2}{4n}\tau^2\geq -r$.  This could lead to a difficulty, pointed
out by N. O'Murchadha, on an asymptotically euclidean manifold where $r$ would be too
negative.  Indeed it is known that in the case $p=2, \; \delta =-1$
there exists a constant $C_P$ such that the following Poincar\'{e} estimate gives an upper bound of
$|\tau|$ in terms of $|D\tau|$:
\begin{equation}
\|\tau\|_{H_{0,-1}}\leq C_P \|D\tau \|_{H_{0,0}} \; .
\end{equation}

If we suppose that an analogous inequality holds for $p>n$, $\delta>-\frac{n}{p}$, {\it i.e},
that there exists a constant $C_P$ such that
\begin{equation}
\| \tau\|_{W^p_{1,\delta}}\leq C_P \|D\tau \|_{W^p_{0,\delta+1}} \; ,
\end{equation}
then by using the Sobolev embedding theorem:
\begin{equation}
\sup_M\sigma^{2\alpha}\tau^2 \leq CC^2_p \|D\tau \|^2_{W^p_{0,\delta+1}} \; , \quad
\mbox{for all } \alpha<\delta+\frac{n}{p} \; .
\end{equation}
This inequality implies that the condition $b\geq -r$ can be satisfied only if $r$ satisfies
the condition
\begin{equation}
-r\sigma^{2\alpha}\leq \frac{4n}{n-2}CC^2_P \epsilon \; .
\end{equation}
We can estimate the value of $\epsilon$ as follows, considering for simplicity the vacuum
case $q=0=v=0$.  The supersolution $\Phi_\mu$ satisfies the equation
\begin{equation}
\triangle_\gamma \Phi_\mu = -A \Phi_\mu^{-P}
\end{equation}
We know that $\Phi_\mu \geq 1$; therefore $A\Phi^{-P}_\mu \leq A$ and $\Phi_\mu \leq \Psi_\mu$,
where $\Psi_\mu$ is the solution with $\Psi_\mu -1\in W^p_{s+2,\delta}$ of the equation
\begin{equation}
\triangle_\gamma \Psi_\mu = -A \equiv -\mu / \sigma^\alpha \; .
\end{equation}
Obviously $\Psi_\mu = 1+\mu w_1$, where $w_1$ depends only on $(M,\gamma)$, satisfies the
equation
\begin{equation}
\triangle_\gamma w_1 = -1/ \sigma^\alpha
\end{equation}
and tends to zero at infinity.
The inequality to satisfy is then 
\begin{equation}
\|D\tau \|^2_{W^p_{0,\delta+2}}\leq (C\mu - S)(1+C_1\mu)^{-4n/(n-2)} \; , \quad \mbox{with }
C_1 =\sup_M w_1 \; .
\end{equation}
The right hand side is maximum for a finite value $\mu = \mu_0$ with $\mu_0$ given by 
\begin{equation}
\mu_0 = \frac{(n-2)C+4nC_1S}{(3n+2)CC_1} \; .
\end{equation}
(Note that $\mu_0 > \frac{S}{C}$ if $n>2$.)  We find therefore
\begin{equation}
\epsilon = \left\{\frac{4n}{n+2}\left(1+\frac{C_1S}{C}\right)\right\}^{-4n/(n-2)}
\frac{n-2}{3n+2}\left(S+\frac{C}{C_1}\right) \; .
\end{equation}
It is an open problem to prove the analogue of the Poincar\'{e} inequality in $W^p_{0,\delta}$
and to decide whether the restriction imposed on $r$ implies that $(M,\gamma)$ is in the 
positive Yamabe class or not.  The conclusion may be (cf.~\cite{O'Murchadha1988}) related to 
the existence of apparent horizons as in the proof by Schoen and Yau \cite{SchoenandYau} of the
positive energy theorem.

\section{Coupled Constraints With Unscaled \\
	Sources}

We treat in this section the system of constraints for unscaled sources on a maximal submanifold 
in the case $n=3$.  This system is coupled if the given initial momentum $J$ does not vanish.  The
equations are:
\begin{eqnarray}
&\triangle_\gamma \varphi - r\varphi + a(\beta)\varphi^{-7} + c\varphi^5 = 0& \nonumber\\ 
&\nabla_i\left\{(2N)^{-1}({\cal L}\beta)^{ij}\right\} = \nabla_i \left\{(2N)^{-1}u^{ij}\right\}
+\varphi^{10}J^j& .
\end{eqnarray}
The functions $c\geq 0$ and $N>0$ and the tensor $u$ and vector $J$ are given on $(M,\gamma)$.
We denote by $\beta_0$ the solution of the equation:
\begin{equation}
\nabla_i\{(2N)^{-1}({\cal L}\beta_0)^{ij}\} = \nabla_i\{(2N)^{-1}u^{ij}\} \; .
\end{equation}
We have the following straightforward result.

{\bf Theorem.}  We suppose that the given quantities $r, \; c, \; J\in W^p_{s,\delta+2}, \; 
u\in W^p_{s+1,\delta+1}, \; 1-N\in W^p_{s+2,\delta}, \; p>n, \; \delta>-n/p$ are such that there
exist positive functions $A_-, \; A_+ \in W^p_{s,\delta+2}$ with:
\begin{equation}
A_-<a(\beta_0)<A_+
\end{equation}
and such that the equation:
\begin{equation}
\triangle_\gamma \varphi - r\varphi +A_+\varphi^{-7} + c\varphi^5 = 0
\end{equation}
admits as supersolution the constant $m_{A_+}\geq 1$, and that the analogous equation constructed
with $A_-$ admits as subsolution the constant $\ell_{A_-}\leq 1$, $\ell_{A_-}>0$.  Then there
exists $\epsilon >0$ such that
\begin{equation}
\|J\|_{W^p_{0,\delta+2}}\leq \epsilon
\end{equation}
implies that the coupled constraint equations have a solution $(\beta,\varphi)$ with $\varphi >0$
and $\beta, \; 1-\varphi \in W^p_{s+2,\delta}$.

{\it Proof.}  It is elementary to check that $\ell_{A_-}$ and $m_{A_+}$ are admissible sub and
supersolutions of the Lichnerowicz equation constructed with $a(\beta)$ if
\begin{equation}
A_-\leq a(\beta)\leq A_+ \; .
\end{equation}
In this case the equation has a solution $\varphi$ with $\ell_{A_-} \leq  \varphi \leq m_{A_+}$.

The momentum current being independent of other quantities its bound
does not affect other estimates.

We will construct a sequence $\varphi_\nu , \beta_\nu$ as in the previous section.  We 
now have to show $A_-<a(\beta_{\nu-1})<A_+$ implies the same inequalities for $a(\beta_\nu)$
if $\varphi_{\nu-1}\leq m_{A_+}$ and $\|J\|_{W^p_{s,\delta+2}}$ is small enough.  We use the
fact that (the notation $| \; |$ means here the $\gamma$ norm):
\begin{eqnarray}
|\{a(\beta_0)\}^{1/2} &-& (16N)^{-1}| {\cal L}(\beta - \beta_0)\| \nonumber\\
&\leq& \{a(\beta)\}^{1/2} \leq \{a(\beta_0)\}^{1/2}
+ (16N)^{-1} | {\cal L}(\beta - \beta_0) | \; .
\end{eqnarray}
We deduce from the momentum constraint satisfied by $\beta_\nu$ the elliptic estimate
\begin{equation}
\|\beta_\nu - \beta_0 \|_{W^p_{2,\delta}} \leq Cm^{10}_{A_+} \|J\|_{W^p_{0,\delta+2}} \; .
\end{equation}
The proof is then completed along the same lines of previous proofs.
 
\section{Acknowledgements}
The authors thank Niall O'Murchadha for helpful remarks, and Sarah and Mark Rupright for
technical assistance in preparing the manuscript.  J.I. received support from National
Science Foundation Grant Number (PHY 98-00732) and J.W.Y. from National Science Foundation
Grant Number PHY 94-13207.  J.W.Y.  thanks the group Gravitation and Cosmologie
Relativistes, Universit\'{e} de Paris VI for hospitality while part of this paper was
written. The research was supported in part under grant number
PHY 94-07194 at the Institute for Theoretical Physics
at the University of California at Santa-Barbara.

\section{Appendix A: Elliptic linear systems on manifolds Euclidean
	at infinity}

For the convenience of the reader we recapitulate here some known
facts. 

A linear differential operator of order $m$ from sections $u$
of a tensor bundle $E$ over a smooth Riemannian manifold $(M,\gamma)$
into sections of another such bundle $F$ reads
\begin{equation}
L u \equiv \sum^m_{k=0} a_k D^k u
\end{equation}
with $a_k$ a linear map from tensor fields to tensor fields 
given also by tensor fields over $M$.

The principal symbol of the operator $L$ at a point
$x \in M$, for a covector $\xi$ at $x$, is the linear
map from $E_x$ to $F_x$ determined by the contraction
of $a_m$ with $(\otimes \xi)^m$. The operator is said
to be elliptic if for each $x\in M$ and $\xi \in T^*_x M$
its principal symbol is an isomorphism from $E_x$ onto
$F_x$ for all $\xi \neq 0$.

\underline{Example.} The conformal Laplace operator
in a metric $\gamma$ on $M$ acting from vector fields
$\beta$ into vector fields is
\begin{equation}
\nabla_i ({\cal L} \beta)^{i j} \equiv \nabla_i 
	\left( \nabla^i \beta^j + \nabla^j \beta^i
	- \frac{2}{n} \gamma^{i j} \nabla_k \beta^k \right) \;.
\end{equation}
Its principal symbol at $x$, with $\xi \in T_x M$, is
the linear mapping from covariant vectors $b$ into
covariant vectors $a$ given by
\begin{equation}
\xi^i \xi_i b_j + \xi^i \xi_j b_i - \frac{2}{n} \xi_j \xi^k b_k = a_j \;.
\end{equation}
This linear mapping is an isomorphism if $\xi \neq 0$ because
\begin{equation}
(\xi^i \xi_i) (b^j b_j) + \left(1-\frac{2}{n}\right) (\xi^i b_i)^2 > 0
\end{equation}
if $\xi \neq 0$ and $b \neq 0$. The conformal Killing
operator is elliptic.

{\bf Theorem.} Let $(M,e)$ be a (complete) Riemannian manifold 
Euclidean at infinity. Let
\begin{equation}
L u \equiv \sum_{k=0}^{m} a_k D^k u
\end{equation}
be an elliptic operator on $(M,e)$. Suppose the coefficients
of $L$ satisfy the following hypotheses
\begin{enumerate}
\item	There is a $C^\infty$ tensor field $A_m$ on $M$, constant in each
	end of $(M,e)$ such that for some $p$ with $1 < p < +\infty$
	\[ 
	a_m - A_m \; \in W^p_{s_m, \delta_m}\;,\;\; 
	s_m > \frac{n}{p} + 1 \;,\;\; 
	\delta_m > -\frac{n}{p} \;, 
	\]
\item 	\[
	a_k \in W^p_{s_k, \delta_k}, s_k > \frac{n}{p} + k =m+1,
	\delta_k > m - k - \frac{n}{p}, 0 \leq k.
	\]
\end{enumerate}
Then for each $s$ such that $s_k + m \geq s \geq m$ the operator
$L$ maps $W^p_{s,\delta}$ into $W^p_{s-m,\delta+m}$ with
finite dimensional kernel and closed range if
\begin{equation}
-\frac{n}{p} < \delta < -m + n - \frac{n}{p} \; .
\end{equation}
If, moreover, $L$ is injective on $W^p_{s,\delta}$ then it is 
an isomorphism and there is a number $C$ such that for each
$u$ in $W^-_{0,\delta}$ the following inequality holds:
\begin{equation}
\| u \|_{W^p_{s,\delta}} \leq C \|L u\|_{W^p_{s-m, \delta+m}} \; .
\end{equation}
This theorem applies to the Poisson operator $\triangle - k$
under the hypothesis indicated in the theorem in Appendix B.

\section{Appendix B:  Solution of $\triangle_\gamma \varphi \equiv f(x,\varphi)$ on an 
Asymptotically Euclidean $(M,\gamma)$}

Let $\triangle_\gamma$ denote the Laplace operator on scalar functions on $(M,\gamma)$.  Let
$f$ be a real valued function on $M\times I$, with $I$ an interval of $R$, given by
$(x,y)\mapsto f(x,y)$.  We will show that the sub and supersolution method used by one
of us (J.I.) in the case of a compact manifold can be extended to asymptotically
euclidean ones.  Recall that $(M,\gamma)$ is a $(p,\sigma,\rho)$ asymptotically euclidean
manifold $M$ of dimension n, if $\gamma - e\in W^p_{\sigma,\rho}$ with $(M,e)$ euclidean
at infinity and $\rho > -\frac{n}{p}$, $\sigma > \frac{n}{p} + 1$.

{\bf 1.  Linear Equations}

{\it Definition.}  Suppose $M$ is not compact.  We say that $f$ tends to a value $c\in {\mathbb R}$
at infinity if for any $\epsilon\geq 0$ there exists a compact $S$ such that:
\begin{equation}
\sup_{M-S}|f-c|\leq \epsilon \; .
\end{equation}

{\bf Lemma 1} (maximum principle).  Let $(M,\gamma)$ be an asymptotically euclidean
manifold.  Suppose that a $C^2$ function $\varphi$ on $M$ satisfies an inequality of the form:
\begin{equation}
\triangle_\gamma \varphi + a.D\varphi - h\varphi \leq 0
\end{equation}
with a dot denoting the scalar product in the metric $\gamma$, while $a$ and $h$ are
respectively a vector field and a function on $M$, both bounded.  Suppose that $h\geq 0$ on
$M$.  Then

a.  If $\varphi$ tends to $c>0$ at infinity then there exists a number $\ell>0$ such that
$\varphi \geq \ell$ on $M$.

b.  If $\varphi$ tends to $c=0$ at infinity then $\varphi \geq 0$ on $M$.

{\it Proof.}  One knows by the classical maximum principle that if $\varphi$ attains
a nonpositive minimum $\lambda$ at a point of $M$ then $\varphi \equiv \lambda$ on $M$.  
Also, if $D$ is a bounded domain of $M$ with smooth boundary $\partial D$ and if the function
$\varphi$ attains a nonpositive minimum in $D\cup \partial D$ this minimum must be attained
on the boundary $\partial D$.

a.  Choose $\epsilon\geq 0$ so small that $\epsilon <c$. If $\varphi$ tends to $c$ at infinity
there is a compact $S$ such that $\varphi \geq c-\epsilon >0$ on $M-K$.  Imbed $S$ in a 
relatively compact domain $D$ with smooth boundary $\partial D$.  On $\partial D$, $\varphi$
takes positive values, therefore $\varphi$ does not attain a nonpositive minimum on the compact
set $D\cup \partial D$;  it attains a positive minimum $c^\prime$.  The number $\ell$ is the 
smaller of the two positive numbers $c-\epsilon$ and $c^\prime$.

b.  Suppose that $\varphi$ takes a negative value $\alpha$ on $M$.  Choose $\epsilon < |\alpha|$.  
There is a compact $S$ such that
\begin{equation}
\sup_{M-S}|f|\leq \epsilon \; .
\end{equation}
Take a relatively compact open set $D$ containing $S$.  If $\varphi$ takes a nonpositive
minimum it is on the boundary $\partial D$, {\it i.e.}, in $M-S$, which contradicts the fact that
the absolute value of this minimum is necessarily greater than or equal to $|\alpha |$, itself 
greater than $\epsilon$, which is the maximum of $|\varphi |$ in $M-S$.

{\bf Theorem.}  Let $(M,\gamma)$ be an $(p,\sigma,\rho)$ asymptotically euclidean manifold.
Let $k\in W^p_{s,\delta+2}$, $\delta > -\frac{n}{p}$ be given.  The
operator $\triangle_\gamma - k$ is injective on $W^p_{s+2,\delta}$
if either
\begin{eqnarray}
&1.& \quad k \geq 0 \; , \; s > \frac{n}{p} \;.\\
&2.& \quad \int_M \{|Df|^2 + kf^2 \}\mu_\gamma > 0 \quad \mbox{for all  }f\in C^\infty_0, \; \; 
f \not\equiv 0 \; ,\nonumber\\
& & \quad s\geq 0 \; \mbox{and }\delta >\frac{n}{2}-\frac{n}{p}-1 \; 
\mbox{if } p\neq2, \; \delta\geq -1 \mbox{ if } p=2 \; . \nonumber\\
\end{eqnarray}

{\bf Corollary.}  Under the hypotheses 1 or 2 the operator $\triangle_\gamma -k$ is an isomorphism
from $W^p_{s+2,\delta}$ onto $W^p_{s,\delta+2}$ if $s\leq \sigma -1- \frac{n}{p}, \; -\frac{n}{p}<
\delta < -\frac{n}{p}
+n-2$.

{\it Proof.}  1.  If $s>\frac{n}{p}$, a solution in $W^p_{s+2,\delta}$ with $\delta>-\frac{n}{p}$ 
is in $C^2_\alpha$ for some positive $\alpha$.  The difference 
$\gamma -e$ is in $C^1_\beta$ for some positive $\beta$.  The maximum principle applies and
shows that $u\equiv 0$ on $M$.

2.  The solution $u\in W^p_{2,\delta}$ is not necessarily $C^2$.  To prove that $u\equiv 0$ we will 
multiply by $u$ the equation and integrate on $M$.

If $u\in C^\infty_0$, then
\begin{equation}
\int_M u\triangle_\gamma u\mu_\gamma = -\int_MDu.Du\mu_\gamma \; .
\end{equation}
We can estimate the integrals in the above formula in terms of the Sobolev norm.  In the case
$p=2$ we have
\begin{equation}
\int_M u\triangle_\gamma u\mu_\gamma \leq \|u \|_{H_{0,\delta}}
\| \triangle_\gamma u \|_{H_{0,\delta+2}}\sup_M (1+d^2)^{-(\delta+1)} \; ,
\end{equation}
which is a bounded quantity whenever $\delta+1\geq 0$.

In the case $p\neq 2$ we have
\begin{equation}
\int_M u\triangle_\gamma u\mu_\gamma \leq \|u\|_{W^p_{0,\delta}} 
\| \triangle_\gamma u \|_{W^p_{0,\delta+2}} \| (1+d^2)^{-(\delta+1)} \|_{L^{p^\prime}} \; , \quad
p^\prime = \frac{p}{p-2} \; .
\end{equation}
The considered $L^{p^\prime}$ norm is bounded if
\begin{equation}
2p^\prime (\delta +1)>n \; , \quad \mbox{\it i.e. , } \delta+1>\frac{n}{2}-\frac{n}{p} \;.
\label{Eq:BoundCondition}
\end{equation}
The same kind of estimate
applies to the integral of $Du.Du$.

The density of $C^\infty_0$ in the weighted Sobolev spaces shows then that a solution
$u\in W^p_{2,\delta}$ satisfies the equality
\begin{equation}
\int_M u\triangle_\gamma u\mu_\gamma = -\int_M Du.Du\mu_\gamma = \int_M ku^2\mu_\gamma \; .
\end{equation}
Therefore, under the hypothesis made, we have $Du=0$; hence 
$u=$constant and $u=0$ because $u$ tends to zero at infinity.

The corollary is a consequence of the general theorem on elliptic systems on an asymptotically
euclidean manifold recalled in Appendix A.

If $n>2$ the inequality $\delta < -\frac{n}{p} + n - 2$ is
compatible with (\ref{Eq:BoundCondition}) if $p \neq 2$
[respectively with $\delta \geq -1$ if $p=2$].

{\it Remark.} Under the hypothesies made on $(M,\gamma)$ and $k$,
the solution $u \in W^p_{s+2,\delta}$ of an equation
\begin{equation}
\triangle_\gamma u -k u = v
\end{equation}
with $v\in W^p_{s,\delta^\prime + 2}$ for some $\delta^\prime$
such that $\delta \leq \delta^\prime < -\frac{n}{p} + n-2$
is in $W^p_{s+2,\delta^\prime}$ if $p>\frac{n}{2}$.
Indeed $u \in W^p_{s+2,\delta}$ and $k \in W^p_{s,\delta+2}$
imply that $k u \in W^p_{s,\delta^{\prime \prime}+2}$,
since $s < 2 s + 2 - \frac{n}{p}$ if $p > \frac{n}{2}$, 
$\delta^{\prime \prime} < \delta + (\delta + \frac{n}{p})$.
Since $u$ satisfies
\begin{equation}
\triangle_\gamma u = k u + v \;\; \in W^p_{s, \;
	\inf (\delta^{\prime \prime}, \delta^{\prime}) + 2} \; ,
\end{equation}
we have
\begin{equation}
u \in W^p_{s+2,\; \inf(\delta^{\prime \prime}, \delta^{\prime})} \; .
\end{equation}
An induction argument shows that $u \in W^p_{s+2,\delta^\prime}$.

{\bf 2. Non-Linear Equations}

We suppose that the function $f$ is smooth in $y$ and $W^p_{s,\beta+2}$ in $x$.  To make it
more transparent we take $f$ as a finite sum of products of functions of $x$ by functions of $y$,
as it appears in the hamiltonian constraint:
\begin{equation}
f(x,y)\equiv \sum^Q_{P=0}a_P(x)b_P(y) \; .
\end{equation}

We make the following hypothesis:

{\it Hypothesis $H(W^p_{s,\delta})$.}

\begin{enumerate}
\item There exists an interval $I\subset\mathbb R$ such that the $b$'s are smooth functions
of $y\in I$.
\item The $a$'s are functions on $M$ in $W^p_{s,\delta+2}$.
\end{enumerate}

{\bf Lemma 1.}  Under the hypothesis $H(W^p_{s,\delta})$ the function on $M$ given by
$x\mapsto f(x,\varphi(x))$, denoted in the sequel $f(x,\varphi)$, has the following properties
when $\varphi$ is continuous and takes its values in a closed interval $[\ell,m]\subset I$:
\begin{enumerate}
\item $f(x,\varphi)\in W^p_{0,\delta+2} \quad \mbox{if } s\geq 0 \; .$
\item If $s>\frac{n}{p} \; , \; \delta>-\frac{n}{p}$ and $D\varphi \in 
W^p_{s^\prime-1, \delta^\prime+1}$ with $\delta^\prime>-\frac{n}{p} \; , \; s\geq s^\prime>
\frac{n}{p}$ then $f(x,\varphi)\in W^p_{s^\prime,\delta+2}$ .
\end{enumerate}

{\it Proof.} Part 1 is trivial.  To prove part 2 one uses the calculus derivation formulas
and the multiplication properties of weighted Sobolev spaces.

{\it Definitions.}  A $C^2$ function $\varphi_-$ on $M$ is called a subsolution of
$\triangle_\gamma\varphi = f(x,\varphi)$ if it is such that on $M$,
\begin{equation}
\triangle_\gamma \varphi_- \geq f(x,\varphi_-) \; .
\end{equation}
A $C^2$ function $\varphi_+$ is called a supersolution if on $M$
\begin{equation}
\triangle_\gamma \varphi_+ \leq f(x,\varphi_+) \; .
\end{equation}

{\bf Theorem 1} (existence).  Let $(M,\gamma)$ be a $(p,\sigma,\rho)$ asymptotically
euclidean manifold $\sigma >\frac{n}{p} + 2$ and $f(x,y)$ a function satisfying the
hypothesis $(H)$ with $s>\frac{n}{p}$ ,  $\delta>-\frac{n}{p}$  .  Suppose the equation
$\triangle_\gamma \varphi = f(x,\varphi)$ admits a subsolution $\varphi_-$ and a supersolution
$\varphi_+$ such that on $M$:
\begin{equation}
\ell \leq \varphi_- \leq \varphi_+ \leq m \; , \quad [\ell,m]\subset I
\end{equation}
and
\begin{equation}
\lim_\infty \varphi_- \leq 1 \; , \quad \lim_\infty \varphi_+\geq 1 \; .
\end{equation}
Suppose that $D\varphi_-\; , \; D\varphi_+ \in W^p_{s^\prime-1,\delta^\prime+1} \; , \quad
s^\prime\geq s \; , \; \delta^\prime >-\frac{n}{p}$ .
Then the equation admits a solution $\varphi$ such that
\begin{equation}
\varphi_- \leq \varphi \leq \varphi_+ \; , \; \; 1-\varphi \in W^p_{s+2,\delta} \mbox{  with }
\delta\leq \beta \mbox{  and  } -\frac{n}{p}<\delta<n-2-\frac{n}{p} \; .
\end{equation}

{\it Proof.}  We construct a solution by induction, starting from $\varphi_-$ .

Let $k$ be a positive function on $M$ such that $k\in W^p_{s,\delta+2}$ and at each point
$x\in M$
\begin{equation}
k(x)\geq \sup_{\ell \leq y \leq m} f^\prime_y(x,y) \; .
\end{equation}
Such a function exists by the hypothesis made on $f$.

We set $\varphi_1 = 1+u_1$.  The linear elliptic equation for $u_1$:
\begin{equation}
\triangle_\gamma u_1 - ku_1 = f(x,\varphi_-) - k(\varphi_- - 1)
\end{equation}
has one solution $u_1\in W^p_{s+2,\delta}\subset C^2$, since the right hand side is in
$W^p_{s,\delta+2}$ and the operator on the left is injective from $W^p_{s+2,\delta}$
into $W^p_{s,\delta+2}$ under the hypothesis made on $s$ and $\delta$.  The function $\varphi_1$
tends to 1 at infinity.

We deduce from the equality and the inequalities satisfied respectively by $\varphi_1$ and
$\varphi_-$ the following inequality:
\begin{equation}
\triangle_\gamma(\varphi_1 - \varphi_-) - k(\varphi_1 - \varphi_-) \leq 0 \; \; ;
\end{equation}
hence, by the maximum principle lemma, since $\varphi_1 - \varphi_-$ tends to $c\geq 0$
at infinity,
\begin{equation}
\varphi_1 \geq \varphi_- \quad \mbox{on } M \; .
\end{equation}
Also,
\begin{equation}
\triangle_\gamma (\varphi_+ - \varphi_1) - k(\varphi_+ - \varphi_1) \leq f(x,\varphi_+)
-f(x,\varphi_-) - k(\varphi_+ - \varphi_-)\; ,
\end{equation}
and
\begin{equation}
f(x,\varphi_+) - f(x,\varphi_-) = (\varphi_+ - \varphi_-) \int^1_0 f^\prime_y(x,\varphi_- +
t(\varphi_+ - \varphi_-))dt \; .
\end{equation}
By the hypothesis made on $k, \; \varphi_+$, and $\varphi_1$ we have on $M$
\begin{equation}
\triangle_\gamma (\varphi_+ - \varphi_1)- k(\varphi_+ - \varphi_1) \leq 0 \; \; ,
\end{equation}
hence
\begin{equation}
\varphi_1 \leq \varphi_+ \; .
\end{equation}
The induction formula is, with $\varphi_n = 1+u_n$:
\begin{equation}
\triangle_g u_n - ku_n = f(x,\varphi_{n-1}) - ku_{n-1} \; .
\label{Eq:InductionFormula}
\end{equation}
We suppose that $\varphi_p$ exists for $0\leq p\leq n-1$ with $\varphi_0 = \varphi_-$ and
$u_p \in W^p_{s+2,\delta}$ for $1\leq p\leq n-1$ and that for these $p$'s
\begin{equation}
\varphi_- \leq \varphi_{p-1} \leq \varphi_p \leq \varphi_+ \; .
\end{equation}
The elliptic theory shows $u_n \in W^p_{s+2,\delta}$ exists.  The functions $\varphi_n$ are
continuous, even $C^2$ since $s>\frac{n}{p}$, and tend to 1 at infinity.  The equality resulting
from the equations satisfied by $\varphi_p$ when $p \leq n-1$ gives
\begin{eqnarray}
\triangle_g \varphi_{n-1} - k\varphi_{n-1} = f(x,\varphi_{n-2}) - k\varphi_{n-2} \; , \\
\nonumber \\
\triangle_g \varphi_p - k\varphi_p = f(x,\varphi_{p-1}) - k\varphi_{p-1} \; .
\end{eqnarray}
One deduces then from the maximum principle lemma that on $M$
\begin{equation}
\varphi_{n-1} \leq \varphi_n \; .
\end{equation}
Analogously one uses the maximum principle and the inequality deduced from the equation
and inequality satisfied by $\varphi_n$ and $\varphi_+$,
\begin{equation}
\triangle_\gamma(\varphi_n - \varphi_+) - k(\varphi_n - \varphi_+)\geq f(x,\varphi_{n-1}) - 
f(x,\varphi_+) - k(\varphi_{n-1}-\varphi_+) \; ,
\end{equation}
to show that $\varphi_{n-1} \leq \varphi_+$ implies $\varphi_n \leq \varphi_+$ .

The sequence of continuous functions $\varphi_n$ has been shown to be pointwise increasing
and bounded.  It is therefore converging at each point $x\in M$ to a limit
$\varphi(x)=1+u(x)$, with $\varphi_- \leq \varphi \leq \varphi_+$ .

To show that $\varphi$ is a solution of the given equation and $\varphi - 1 \in W^p_{s+2,\delta}$
 we proceed as follows.  Since the $\varphi_n$ are continuous and take their values in the interval
$[\ell,m]$, the functions $f(x,\varphi_n) - ku_n$ belong to $W^p_{0,\delta+2}$ with uniformly
bounded norms.  The linear elliptic inequality
(following from (\ref{Eq:InductionFormula}))
\begin{equation}
\|u_{n+1}\|_{W^p_{2,\delta}}\leq C \|f(x,\varphi_n) - ku_n)\|_{W^p_{0,\delta+2}}
\end{equation}
shows that the sequence $u_n$ is uniformly bounded in the $W^p_{2,\delta}$ norm.  Since 
$W^p_{2,\delta}$ is compactly embedded in $W^p_{1,\delta^\prime}$ for any $\delta^\prime
<\delta$, there is a subsequence, still denoted $u_n$, which converges in $W^p_{1,\delta^\prime}$
norm to a function $u\in W^p_{2,\delta}$ .

The functions $f(x,\varphi_n)$ converge to $f(x,\varphi)$ in the $W^p_{1,\delta^\prime+2}$ norm
because of the inequality, which is satisfied if $s>\frac{n}{p} \;, \; \; \delta^\prime>-\frac{n}{p}$ ,
\begin{equation}
\|f(x,\varphi)-f(x,\varphi_n)\|_{W^p_{1,\delta+2}}\leq C\| \varphi - \varphi_n 
\|_{W^p_{1,\delta^\prime}} \|F_1 \|_{W^p_{s,\delta+2}} \; ,
\end{equation}
where $C$ depends only on $(M,e)$ and $F_1 \in W^p_{s,\delta+2}$ is a function on $M$, which
exists by the hypothesis on $f$, such that
\begin{equation}
F_1(x)\geq \sup_{y\in [\ell,m]}|f^\prime_y(x,y)| \; .
\end{equation}
These convergences imply that the limit $\varphi =1+u$ satisfies the equation in a
generalized sense.  From the linear theory, we find that the equation satisfied by $u$ and the
fact that (cf. above lemma) that $f(x,\varphi)\in W^p_{1,\delta+2}$ (also $\in W^p_{2,\delta+2}$
since $u\in W^p_{2,\delta}$) that $u\in W^p_{3,\delta}$ .  An induction argument completes the
proof that $u\in W^p_{s+2,\delta}$ .

The theorem holds with $s=0$ if $f$ is an increasing function of $y$.  An example is treated in
Section \ref{DiscontinuousSources}.

For references before 1980, see \cite{cbyo1980}

\end{document}